\documentclass[12pt]{article}

\usepackage[utf8]{inputenc}
\usepackage{latexsym, graphicx} 
\usepackage{amsmath,amsfonts,amssymb}
\usepackage{slashed}
\usepackage{cancel}
\usepackage{mathrsfs,dsfont}
\usepackage{tensor}
\usepackage{xcolor}
\usepackage{cite}
\usepackage[colorlinks=true,linkcolor=blue,citecolor=blue,urlcolor=blue,filecolor=black]{hyperref}

\renewenvironment{thebibliography}[1]{
	\begin{oldthebibliography}{#1}
		\setlength{\itemsep}{2pt}
		\setlength{\parskip}{2pt}
	}
	{
	\end{oldthebibliography}
}

\numberwithin{equation}{section} 

\newcommand*{\dd}{\mathop{}\!d}

\newcommand*{\x}{\mathbf{x}}

\newcommand*{\HH}{\mathbb{H}}

\newcommand{\E}{{\rm e}}

\newcommand{\p}{\hat{p}}

\begin{document}
	
	\begin{titlepage}
		\thispagestyle{empty}
		
		\begin{flushright}
		\end{flushright}
		
		\vskip1cm
		
		\begin{center}  
			{\Large\textbf{Celestial decomposition of Wigner's particles}}
			
			\vskip1cm
			
			\centerline{Lorenzo Iacobacci and Kevin Nguyen}
			
			\vskip1cm
			
			{\it{Universit\'e Libre de Bruxelles and International Solvay Institutes,\\ ULB-Campus Plaine
					CP231, 1050 Brussels, Belgium}}\\
			\vskip 1cm
			{lorenzoiacobacci@gmail.com, kevin.nguyen2@ulb.be}
			
		\end{center}
		
		\vskip1cm
		
		\begin{abstract} 
			We provide a detailed decomposition of Wigner's particles, defined as unitary irreducible representations of the Poincar\'e group, in terms of unitary representations of its Lorentz subgroup. As pointed out before us, this decomposition only involves Lorentz representations belonging to the principal continuous series, and further underpins the connection between scattering amplitudes and conformal correlation functions discussed in the context of celestial holography. We provide very explicit formulae for the decomposition of particles of arbitrary mass and (half-)integer spin and for any spacetime dimension. We emphasise that this decomposition is unique and comes equipped with a specific inner product on the corresponding Hilbert space. We also confirm that unitary translations mix Lorentz representations within the principal continuous series only, as required. Implications to the celestial holography program are indicated.  
		\end{abstract}
		
	\end{titlepage}
	
	{\hypersetup{linkcolor=black}
		\tableofcontents 
	}
	
	\section{Introduction}
	Unitary representations of symmetry groups play a fundamental role in quantum mechanics and quantum field theory (QFT). Observers related to one another by a continuous symmetry should measure the same transition probabilities, which implies that the latter are represented on the Hilbert space of physical states by unitary operators \cite{Wigner1931,Weinberg:1995mt}. The physical Hilbert space is the carrier space of a unitary representation of the symmetry group. Unitary irreducible representations (UIRs) can then be used as the buildings blocks that can potentially  appear within a specific theory.\footnote{Beware of reducible but indecomposable representations, however.} Most notably this methodology led Wigner to propose a classification of stable particles in terms of UIRs of the inhomogeneous Lorentz group $\operatorname{ISO}(1,d+1)$ (also known as Poincar\'e group), which still constitutes  the basis of our understanding of relativistic particle scattering \cite{Wigner:1939cj}. Another big success of this approach lies in the study of conformal field theories (CFT), whose Hilbert space carries unitary representations of the pseudo-orthogonal group $\operatorname{SO}(2,d)$. In that context group theory is so powerful that any given theory can in principle be solved non-perturbatively \cite{Poland:2018epd}.
	
	In recent years substantial effort has been put in revisiting the theory of particle scattering through the lens of the Lorentz group $\operatorname{SO}(1,d+1) \subset \operatorname{ISO}(1,d+1)$, with the hope of exploiting powerful methods developed in the context of conformal field theory. This program, which goes under the name of \textit{celestial holography}, is reviewed for example in \cite{Pasterski:2021raf,McLoughlin:2022ljp}. It should be emphasised that if one aims at describing a quantum theory where Poincar\'e symmetries are realised unitarily on the physical Hilbert space, then one necessarily needs to consider unitary representations of its Lorentz subgroup. One should therefore be very careful in trying to import methods from conformal field theory, since analytic continuation from $\operatorname{SO}(2,d)$ to $\operatorname{SO}(1,d+1)$ turns unitary representations into non-unitary ones (called reflection-positive instead). At this time, it is still unknown which of the many results from standard conformal field theory can be safely transposed to theories with unitary $\operatorname{SO}(1,d+1)$ symmetry. Existing work on the subject points to even greater tractability in some aspects \cite{Gadde:2017sjg}.
	
	In this paper we intend to work out the decomposition of Wigner particles, defined as UIRs of the Poincar\'e group $\operatorname{ISO}(1,d+1)$, into UIRs of its Lorentz subgroup. While this problem has been studied in the past  \cite{Joos2006ZurDD,Cantoni1975OnTS,Chakrabarti1967ONT,Chakrabarti1968LorentzBO,Mukunda1968ZeroMassRO,Shapiro1962ExpansionOT,Smith1978MatrixEE,Rhl1969TheCO,Weidemann1980QuantumFI,Lomont1964TheRO,MacDowell1972ReductionOT,Zastavenko2009IntegralTO,Kuangchao1958THESI,Zmuidzinas1966UnitaryRO,Steiger1971PoincarIrreducibleTO}, our aim is to present the solution in the language of \textit{elementary representations}, which are carried by basis states $|\Delta,\vec x\rangle_\sigma$ labeled by a fixed scaling dimension $\Delta$, a continuous parameter $\vec x \in \mathbb{R}^d$ and a finite set of spin indices~$\sigma$.\footnote{Elementary representations cover both unitary and reflection-positive representations, depending on the value of $\Delta$.} For $d=2$ which is most relevant to phenomenology, this has first been performed in \cite{Mukunda1968ZeroMassRO} for massless particles of arbitrary integer spin and in \cite{MacDowell1972ReductionOT} for massive scalar particles. The resulting decompositions have then independently re-appeared in the context of celestial holography in a slightly disguised form relating plane waves to \textit{conformal primary wavefunctions} \cite{deBoer:2003vf,Cheung:2016iub,Pasterski:2017kqt,Pasterski:2016qvg}, with extensions to spinning massive wavefunctions \cite{Law:2020tsg,Iacobacci:2020por,Narayanan:2020amh}. While the central decomposition formulae do appear in this second body of work, a complete group theoretic analysis was not provided. In particular, the physical Hilbert space was not explicitly described and questions of unitarity left aside. In this paper we close this gap, providing at the same time the generalisation to arbitrary dimension $d$, mass and (half)-integer spin. 
	
	The decomposition of Wigner's UIRs into elementary representations is essentially a problem in harmonic analysis. Indeed, a generic one-particle state can be characterised by a square integrable momentum wavefunction $\psi(\hat p)$, with $\hat p$ a point on the (forward) null cone $\operatorname{LC}_{d+1}^+\subset \mathbb{M}^{d+2}$ in the massless case, or a point on the negatively curved hyperboloid $\mathbb{H}^{d+1}\subset \mathbb{M}^{d+2}$ in the massive case. Here $\mathbb{M}^{d+2}$ denotes the $(d+2)$-dimensional Minkowski spacetime. Both $\operatorname{LC}_{d+1}^+$ and $\mathbb{H}^{d+1}$ are homogeneous spaces with respect to $\operatorname{SO}(1,d+1)$, i.e., they can be realised as the quotient $\operatorname{SO}(1,d+1)/\operatorname{H}$ for some closed subgroup $\operatorname{H} \in \operatorname{SO}(1,d+1)$, and the decomposition we are looking for amounts to a decomposition of a generic square integrable function $\psi(\hat p)$ over a complete set of harmonic functions carrying irreducible representations of $\operatorname{SO}(1,d+1)$. While this philosophy underlies our approach, the description we provide will be rather practical and we refer the interested reader to \cite{Barut:1986dd,Dobrev:1977qv,Chang:2023ttm} for a detailed discussion of harmonic analysis in relation to group theory.   
	
	The paper is organised as follows. In section~\ref{section 2} we briefly review the elementary representations of the Lorentz group $\operatorname{SO}(1,d+1)$ \cite{Dirac:1945cm,HarishChandra1947InfiniteIR,Bargmann:1946me,gel1947unitary,Dobrev:1977qv}. They can be constructed using the method of induced representations, inducing from a representation $[\Delta,\rho]$ of $\mathbb{R} \times \operatorname{SO}(d)$, with $\Delta$ a scaling dimension and $\rho$ an irrep of $\operatorname{SO}(d)$ acting on the spin degrees of freedom. We pay particular attention to the unitary representations of the \textit{principal continuous series} where $\Delta \in \frac{d}{2}+i\mathbb{R}$, since these are the only ones which will appear in the aforementioned harmonic decomposition. In section~\ref{section 3} we describe the decomposition of a generic massless particle UIR. As is well-known from the works of de~Boer--Solodukhin \cite{deBoer:2003vf} and Pasterski--Shao \cite{Pasterski:2017kqt}, a Mellin transform exchanges the particle energy $\omega$ for the scaling dimension $\Delta$ of the elementary representation, integrated over the continuous principal series. The representation of the little group $\operatorname{SO}(d)$ from which Wigner's UIRs are induced is then readily identified with the representation $\rho$ of the corresponding elementary representation, such that the same decomposition applies irrespective of the particle spin. We explicitly discuss the compatibility of this decomposition with the inner products that the one-particle Hilbert space and the Hilbert spaces of the principal continuous series come equipped with. We show how Wigner's UIRs are turned into Lorentz UIRs, working infinitesimally in the main text, and with finite transformations in appendix~\ref{app:1}. Finally we show that unitary translations $U(a)=e^{-i a^\mu \tilde P_\mu}$ mix the Lorentz UIRs within the principal continuous series, and we give the explicit formula describing this mixing. This contrasts with the widespread folklore that translations require to consider representations off the principal continuous series, which are non-unitary. In section~\ref{section 4} we turn to the decomposition of massive particle UIRs. In the spinless case, the decomposition given by Pasterski--Shao--Strominger \cite{Pasterski:2016qvg} is used to easily reach  analogous conclusions. Adding spin introduces substantial technical complications, however. The origin for this complication is the fact that massive particles carry a representation of the little group $\operatorname{SO}(d+1)$, which needs to be decomposed over the representations $\rho$ of $\operatorname{SO}(d)$ labeling the elementary representations. Fortunately, for integer spin all the required technology has already been developed in a different context in \cite{Costa:2011mg,Costa:2014kfa}. In particular a resolution of the identity on the space of square integrable spinning wavefunctions was provided in \cite{Costa:2014kfa}, which has been used in \cite{Law:2020tsg} to discuss massive conformal primary wavefunctions for $d=2$. For half-integer spin, the corresponding resolution of identity has been provided by one of us in \cite{Iacobacci:2020por}. Using these formulae, we are able to describe in detail the decomposition of massive Wigner UIRs into elementary representations of the principal continuous series, for arbitrary dimension $d$ and (half-)integer spin.       
	
	\section{Unitary representations of the Lorentz group}
	\label{section 2}
	Unitary representations of the Lorentz group $\operatorname{SO}(1,d+1)$ have been classified in \cite{Dirac:1945cm,HarishChandra1947InfiniteIR,Bargmann:1946me,gel1947unitary,Dobrev:1977qv}. For the purposes of this work the relevant unitary representations are those of the so-called \textit{principal continuous series}. We will start by describing them in a language most suitable to the present construction and closely related to that of conformal field theory, mostly following the physicist-friendly review \cite{Sun:2021thf}.
	
	While the Lorentz algebra $\mathfrak{so}(1,d+1)$ is usually presented in terms of the abstract generators $J_{\mu\nu}$ as
	\begin{equation}
	\label{Lorentz algebra}
	\left[J_{\mu\nu},J_{\rho \sigma}\right]=-i\left(\eta_{\mu\rho} J_{\nu \sigma}+\eta_{\nu \sigma}J_{\mu \rho}-\eta_{\mu\sigma} J_{\nu\rho}-\eta_{\nu \rho} J_{\mu \sigma} \right)\,,
	\end{equation}
	with $\eta_{\mu\nu}=\text{diag}(-,+,...,+)$, it is very convenient to distinguish the indices $\mu=0$ and $\mu=d+1$ and define the generators $(P_i,K_i,D)$ through
	\begin{equation}
	\label{change of basis}
	J_{0i}=\frac{K_i+P_i}{2}\,, \qquad J_{(d+1)i}=\frac{K_i-P_i}{2}\,, \qquad J_{0(d+1)}=-D\,,
	\end{equation}
	such that the algebra \eqref{Lorentz algebra} takes the form
	\begin{equation}
	\begin{split}
	\left[D,P_i\right]&=iP_i\,, \qquad \left[D,K_i\right]=-i K_i\,, \qquad \left[K_i,P_j\right]=-2i\left(\delta_{ij} D-J_{ij} \right)\,,\\
	\left[J_{ij},P_k \right]&=-i\left(\delta_{ik} P_j-\delta_{jk}P_i \right)\,, \qquad \left[J_{ij},K_k \right]=-i\left(\delta_{ik} K_j-\delta_{jk}K_i \right)\,,\\
	\left[J_{ij},J_{kl}\right]&=-i\left(\delta_{ik} J_{jl}+\delta_{jl}J_{ik}-\delta_{il}J_{jk}-\delta_{jk}J_{il}\right)\,.
	\end{split}
	\end{equation}
	Indeed the Bruhat decomposition of the Lorentz group is then given in terms of the subgroups
	\begin{equation}
	\label{Bruhat}
	\operatorname{N}=\{e^{i b^i K_i} \}\,, \qquad \operatorname{A}=\{e^{i\lambda D}\}\,, \qquad \operatorname{M}=\{e^{i\omega^{ij} J_{ij}}\}\,, \qquad \tilde{\operatorname{N}}=\{e^{i x^iP_i} \}\,. 
	\end{equation}
	The basis generators $\langle J_{ij},P_i,K_i,D \rangle$ with $i=1,...,d$ are naturally associated with the conformal transformations of $\mathbb{R}^d$ with cartesian coordinates $x^i$, which can be viewed as the homogeneous space
	\begin{equation}
	\mathbb{R}^{d} \cong \tilde{\operatorname{N}} \cong \frac{\operatorname{SO}(1,d+1)}{\operatorname{MAN}}\,.
	\end{equation}
	In this basis the quadratic Casimir operator is then given by
	\begin{equation}
	\mathcal{C}_2^{\mathfrak{so}(1,d+1)}=\frac{1}{2} J_{\mu\nu} J^{\mu\nu}=iD(d+iD)-P_i K^i+\frac{1}{2} J_{ij}J^{ij}\,.
	\end{equation}
	Conformal field representations can then be constructed using the method of induced representations, starting from a representation of the isotropy subgroup $\operatorname{MAN}$ spanned by $\langle J_{ij},K_i,D\rangle$ \cite{Mack:1969rr}. The inducing representation is characterised by
	\begin{equation}
	\label{inducing representation}
	D\, |\Delta \rangle_\sigma =i \Delta\,|\Delta \rangle_\sigma \,, \qquad J_{ij}\, |\Delta \rangle_\sigma=(\Sigma_{ij})\indices{_\sigma^{\sigma'}}|\Delta\rangle_{\sigma'}\,, \qquad K_i\, |\Delta \rangle_\sigma=0\,,
	\end{equation}
	where $\Delta \in \mathbb{C}$ and $(\Sigma_{ij})\indices{_\sigma^{\sigma'}}$ are hermitian spin matrices furnishing a unitary representation of $\text{SO}(d)$. The induced representation is then obtained by `translating' this state,
	\begin{equation}
	\label{conformal state}
	|\Delta, \vec x \rangle_\sigma \equiv e^{i x^i P_i} |\Delta \rangle_\sigma\,.
	\end{equation}
	Using the commutation relations of the generators \eqref{change of basis} together with the defining relation \eqref{inducing representation}, one can easily work out the action of the generators on the state \eqref{conformal state}, resulting in
	\begin{equation}
	\label{induced rep}
	\begin{split}
	P_i\, |\Delta, \vec x \rangle &=-i \partial_i\, |\Delta, \vec x\rangle\,,\\
	J_{ij}\, |\Delta, \vec x\rangle&=-i\left(x_i \partial_j-x_j\partial_i+i \Sigma_{ij} \right) |\Delta, \vec x\rangle\,,\\
	D\, |\Delta, \vec x\rangle&=i\left(\Delta+x^i\partial_i \right)|\Delta, \vec x\rangle\,,\\
	K_i\, |\Delta, \vec x\rangle &=i\left(2x_i \Delta+2x_i x^j \partial_j-x^2 \partial_i+2i x^j \Sigma_{ij} \right)|\Delta, \vec x\rangle\,,
	\end{split}
	\end{equation}
	where spin indices have been suppressed for convenience. This provides an \textit{elementary representation} of the Lorentz group, with value of the quadratic Casimir given by 
	\begin{equation}
	\label{Casimir conformal value}
	\mathcal{C}_2^{\mathfrak{so}(1,d+1)}=\Delta(\Delta-d)+\mathcal{C}_2^{\mathfrak{so}(d)}\,,
	\end{equation}
	where $\mathcal{C}_2^{\mathfrak{so}(d)}$ is the Casimir value associated with the defining $\operatorname{SO}(d)$ representation. For a totally symmetric traceless spin-$\ell$ representation, we have $\mathcal{C}_2^{\mathfrak{so}(d)}=\ell(\ell+d-2)$. We denote by $[\Delta,\ell]$ the corresponding representation. 
	The unitarity of the induced representation \eqref{induced rep} depends on a choice of invariant inner product and on the value of the scaling dimension $\Delta$, to be discussed momentarily. Importantly any admissible irreducible representation can be found within such an elementary representation \cite{Langlands,Knapp1,Knapp2}. Of primary interest to us are the unitary representations of the \textit{principal continuous series} that exist for any scaling dimension 
	\begin{equation}
	\label{principal series}
	\Delta=\frac{d}{2}+i\nu\,, \qquad \nu \in \mathbb{R}\,, 
	\end{equation}
	with Casimir value
	\begin{equation}
	\label{Casimir principal series}
	\mathcal{C}_2^{\mathfrak{so}(1,d+1)}=-(\nu^2+d^2/4)+\mathcal{C}_2^{\mathfrak{so}(d)}=-(\nu^2+d^2/4)+\ell(\ell+d-2)\,,
	\end{equation}
	where in the second equality we restricted to symmetric spin-$\ell$ representation of $\operatorname{SO}(d)$.  
	These constitute irreducible unitary representations.
	Note that $\nu$ and $-\nu$ are associated with the same Casimir value, and there exist an isomorphism between these representations known as `shadow transform'. 
	
	The above discussion was phrased in terms of infinitesimal Lorentz transformations. To obtain the finite form of the transformations, we recall that the Lorentz group $\operatorname{SO}(1,d+1)$ is naturally realised as the group of conformal isometries of $\mathbb{R}^d$, whose action we abstractly denote $\vec x \mapsto \vec x'=\Lambda \vec x$ for any $\vec x \in \mathbb{R}^d$. Explicit expressions can be found for example in \cite{DiFrancesco:1997nk}. Such conformal isometries satisfy
	\begin{equation}
	\label{finite conformal transfo}
	\frac{\partial x'^i}{\partial x^j}=\Omega(x)R\indices{^i_j}(x)\,, \qquad \Omega(x)=\left|\frac{\partial \vec x'}{\partial \vec x}\right|^{\frac{1}{d}}\,,
	\end{equation}
	with $\left|\frac{\partial \vec x'}{\partial \vec x}\right|$ the determinant of the transformation, and $R\indices{^i_j}(x) \in \operatorname{SO}(d)$ an orthogonal matrix with unit determinant. For any element $\Lambda \in \operatorname{SO}(1,d+1)$, the group action \eqref{induced rep} exponentiates to
	\begin{equation}
	\label{Lorentz finite transfo}
	U(\Lambda) |\Delta, \vec x \rangle_\sigma=\Omega(x)^{\Delta}\, D(R(x)^{-1})\indices{_\sigma^{\sigma'}} |\Delta, \vec x\,' \rangle_{\sigma'}\,,
	\end{equation}
	where $D(R(x))$ is the spin-$\ell$ representation of $R(x) \in \operatorname{SO}(d)$.
	
	We will consider a Hilbert space admitting an action of $\operatorname{SO}(1,d+1)$ consisting of the direct integral over the principal continuous series \eqref{principal series} with weight $\rho(\nu)$, and fixed $\operatorname{SO}(d)$ spin for simplicity. The generalisation to the direct sum of Hilbert spaces with different $\operatorname{SO}(d)$ spins is trivial. This Hilbert space can be described as follows. Given a complex wavefunction $\psi_\sigma(\nu,\vec x)$, we associate the state
	\begin{equation}
	\label{psi}
	|\psi\rangle =\sum_\sigma \int_{-\infty}^\infty d\nu\, \rho(\nu) \int d^d\vec x\, \psi_\sigma(\nu,\vec x) |\nu,\vec x \rangle_\sigma\,,
	\end{equation}
	where the sum runs over all spin components,
	together with the inner product 
	\begin{equation}
	\label{Lorentz inner product}
	\langle \phi|\psi\rangle=\sum_\sigma \int_{-\infty}^\infty d\nu\, \rho(\nu) \int d^d\vec x\, \phi_\sigma(\nu,\vec x)^*\, \psi_\sigma(\nu,\vec x)\,.
	\end{equation}
	Up to this point the weight $\rho(\nu)$ is arbitrary, although when considering the decomposition of Poincar\'e representations it will take on specific forms. The Hilbert space furnishing the carrier space for the unitary representation is the set of states $|\psi \rangle$ with normalisable wavefunctions $\psi_\sigma(\nu,\vec x)$, i.e., with $\langle \psi|\psi \rangle < \infty$.\footnote{Wavefunctions which coincide everywhere except on a set of measure zero are identified and correspond to the same state.} As physicists we are happy to write an inner product for the states $|\nu,\vec x \rangle_\sigma$ themselves even though they do not strictly belong to the Hilbert space, namely
	\begin{equation}
	\label{inner product}
	{}_{\sigma'}\langle \nu', \vec y\, |\, \nu, \vec x \rangle_\sigma= \frac{\delta(\nu-\nu') \delta(\vec x-\vec y)}{\rho(\nu)}\, \delta_{\sigma \sigma'} \,, 
	\end{equation}
	which should therefore be understood in the sense of distributions. Note that if instead of a direct integral over the principal series, we wanted to consider a direct sum over a discrete set $\mathbf{S}$ of scaling parameters $\nu$, i.e.,
	\begin{equation}
	|\psi\rangle =\sum_\sigma \sum_{\nu \in \mathbf{S}} \rho_\nu \int d^d\vec x\, \psi_{\sigma,\nu}(\vec x) |\vec x \rangle_{\sigma,\nu}\,,
	\end{equation}
	the Dirac delta $\delta(\nu-\nu')$ in \eqref{inner product} would be replaced by a Kronecker delta $\delta_{\nu\nu'}$. Also note that the square integrable wavefunctions $\psi_\sigma(\nu,\vec x) \in L^2(\mathbb{R}^d)$ for any $\nu \in \mathbb{R}^+$ naturally carry the conjugate representation $[d-\Delta,\ell]=[\frac{d}{2}-i\nu,\ell]$. Indeed using integration by parts, one can show that the same transformation for $|\psi \rangle_\sigma$ as given in \eqref{psi} is obtained either by transforming $|\nu,\vec x \rangle_\sigma$ according to \eqref{induced rep} or $\psi(\nu,\vec x)_\sigma$ according to the conjugate representation. Since the conjugate representations $[\frac{d}{2}+i\nu,\ell]$ and $[\frac{d}{2}-i\nu,\ell]$ are isomorphic, it is equivalent to work with the states $|\nu,\vec x\rangle_\sigma$ or with the wavefunctions $\psi_\sigma(\nu,\vec x)$.
	
	Finally we can naturally associate an operator $O_{\Delta,\sigma}(\vec x)$ to the state $|\Delta,\vec x \rangle_\sigma$ through the definition
	\begin{equation}
	|\Delta,\vec x\rangle_\sigma \equiv O_{\Delta,\sigma}(\vec x) |0\rangle\,,   
	\end{equation}
	with $|0\rangle$ the `vacuum' state corresponding to the trivial representation. The inner product \eqref{inner product} can then be interpreted as a two-point correlation
	\begin{equation}
	\label{2-point functions}
	\langle 0| O_{\frac{d}{2}+i\nu',\sigma'}(\vec y)^\dagger\, O_{\frac{d}{2}+i\nu,\sigma}(\vec x) |0 \rangle=\frac{\delta(\nu-\nu') \delta(\vec x-\vec y)}{\rho(\nu)}\, \delta_{\sigma \sigma'}\,.
	\end{equation}
	By contrast, for the representations of the \textit{complementary series} that exist for any $\Delta \in (0,d)$ at zero spin and for any $\Delta \in (1,d-1)$ at nonzero spin, the inner product and corresponding two-point functions take the form, up to normalisation,
	\begin{equation}
	\label{familiar}
	{}_{\sigma'}\langle \Delta', \vec y\, | \Delta, \vec x \rangle_\sigma=\langle 0| O_{\Delta',\sigma'}(\vec y)^\dagger\, O_{\Delta,\sigma}(\vec x) |0 \rangle= \frac{\delta(\Delta-\Delta')}{|\vec x-\vec y|^{2\Delta}}\, \delta_{\sigma\sigma'}\,.
	\end{equation}
	Again the Dirac delta is to be replaced by a Kronecker delta for a discrete spectrum of scaling dimensions. Even though the two-point functions \eqref{2-point functions}, contrary to \eqref{familiar}, may appear unfamiliar from the perspective of standard conformal field theory which treats of unitary representations of $\operatorname{SO}(2,d)$, they are completely appropriate to the principal continuous series representations of $\operatorname{SO}(1,d+1)$.  
	
	\section{Decomposition of massless particles}
	\label{section 3}
	In this section we study how massless particle states, defined as UIRs of the Poincar\'e group $\operatorname{ISO}(1,d+1)$, decompose into UIRs of its Lorentz subgroup. The resulting Mellin decomposition of the momentum wavefunctions is well-known since the works of de Boer--Solodukhin \cite{deBoer:2003vf} and Pasterski--Shao \cite{Pasterski:2017kqt}, and has appeared even earlier in the mathematics literature \cite{Mukunda1968ZeroMassRO}. See also \cite{Banerjee:2018gce}. Here we revisit this decomposition, putting particular emphasis on the group theoretic aspects.
	
	We recall that massless wavefunctions are defined on the forward null cone $\operatorname{LC}^+_{d+1}$, which is identified with the homogeneous space
	\begin{equation}
	\operatorname{LC}^+_{d+1}\cong \operatorname{\tilde N A}\cong \frac{\operatorname{SO}(1,d+1)}{\operatorname{ISO}(d)}\,.
	\end{equation}
	Following Wigner's method of induced representation, massless particle states are defined starting from the reference null momentum
	\begin{equation}
	k^\mu=\omega_0\, (1,\vec 0,1)\,,
	\end{equation}
	expressed in terms of an arbitrary reference energy scale $\omega_0$,
	together with a UIR of the massless little group $\operatorname{ISO}(d) \subset \operatorname{SO}(1,d+1)$ generated by $\langle J_{ij},K_i \rangle$. We can characterise this inducing represensation by\footnote{We do not consider continuous/infinite-spin representations, where $K_i$ would have non-trivial action.} 
	\begin{align}
	\label{massless little group rep}
	J_{ij}\, |k\rangle_\sigma=(\Sigma_{ij})_\sigma{}^{\sigma'}|k\rangle_{\sigma'}\,, \qquad K_i\, |k \rangle_\sigma=0\,, \qquad \tilde P_\mu\, |k \rangle_\sigma=k_\mu\, |k\rangle_\sigma\,,  
	\end{align}
	where $\Sigma_{ij}$ is again a hermitian matrix furnishing a unitary representation of $\operatorname{SO}(d)$. Note that the operator $\tilde P_\mu$ above is the standard momentum generator, not to be confused with the boost generators $P_i$ defined in \eqref{change of basis}.
	A generic momentum state is then obtained by applying the remaining boost generators,
	\begin{equation}
	\label{generic massless momentum state}
	|p\rangle_\sigma \equiv e^{ix^i P_i}\, e^{i\ln(\omega/\omega_0)\, D}\, |k\rangle_\sigma\,.
	\end{equation}
	The last equation is just one specific way to implement Wigner's induced representation, resulting in the specific momentum parametrisation \cite{Nguyen:2023vfz,Nguyen:2023miw} 
	\begin{equation}
	\label{massless momentum parametrisation}
	p^\mu=\omega q^\mu(\vec x)\,, \qquad q^\mu(\vec x)\equiv (1+x^2,2\vec x,1-x^2)\,.
	\end{equation}
	The action of the Lorentz generators on the generic state \eqref{generic massless momentum state} can be worked out using the algebra relations together with \eqref{massless little group rep}-\eqref{generic massless momentum state}, yielding
	\begin{equation}
	\label{massless rep}
	\begin{split}
	P_i\, |p\rangle &=-i \partial_i\, |p \rangle\,,\\
	J_{ij}\, |p\rangle&=-i\left(x_i \partial_j-x_j\partial_i+i \Sigma_{ij} \right) |p\rangle\,,\\
	D\, |p\rangle&=i\left(-\omega \partial_\omega+x^i\partial_i \right)|p\rangle\,,\\
	K_i\, |p\rangle &=i\left(-2x_i \omega \partial_\omega+2x_i x^j \partial_j-x^2 \partial_i+2i x^j \Sigma_{ij} \right)|p\rangle\,,
	\end{split}
	\end{equation}
	where spin indices have again been suppressed for convenience. The astute reader will have noticed that this is essentially the same as \eqref{induced rep} up to the replacement $\Delta \leftrightarrow -\omega \partial\omega$. Making this step precise will be the key to decomposing a massless particle UIR into Lorentz UIRs. 
	
	A generic state in the corresponding one-particle Hilbert space takes the form
	\begin{equation}
	\label{generic massless state}
	|\psi \rangle=\sum_\sigma \int [d^{d+1}  p(\omega,\vec x)]\, \psi_\sigma(\omega,\vec x) |p(\omega,\vec x)\rangle_\sigma\,,
	\end{equation}
	with $\psi_\sigma(\omega,\vec x)$ any complex wavefunction normalisable with respect to the inner product
	\begin{equation}
	\label{massless inner product}
	\langle \phi | \psi \rangle=\sum_\sigma \int [d^{d+1} p(\omega,\vec x)]\, \phi_\sigma(\omega,\vec x)^*\, \psi_\sigma(\omega,\vec x)\,,    
	\end{equation}
	and with $[d^{d+1} p(\omega,\vec x)]$ the Lorentz-invariant measure on the null cone $\operatorname{LC}^+_{d+1}$,
	\begin{equation}
	[d^{d+1} p(\omega,\vec x)]= \omega^{d-1} \dd\omega \dd^d\vec x\,.
	\end{equation}
	The complex wavefunctions $\psi_\sigma(\omega,\vec x)$ again naturally carry the conjugate unitary representation, isomorphic to the original representation. Thus the Hilbert space is isomorphic to the space of square integrable complex wavefunctions $L^2(\operatorname{LC}^+_{d+1},[d^{d+1} p])$. We note that in the limit of vanishing frequency $\omega$, square integrability of the wavefunctions requires 
	\begin{equation}
	\label{zero frequency behavior}
	\psi_\sigma(\omega,\vec x)=o\left(\omega^{-\frac{d}{2}} \right)\,, \qquad (\omega \to 0)\,.
	\end{equation}
	Additionally, even though the states $|p(\omega,\vec x)\rangle_\sigma$ do not strictly belong to this Hilbert space, one can assign them the distributional inner product 
	\begin{equation}
	{}_{\sigma'}\langle p_1|\,p_2\rangle_\sigma=(\omega_1)^{1-d}\, \delta(\omega_1-\omega_2) \delta(\vec x_1-\vec x_2) \delta_{\sigma\sigma'}\,,
	\end{equation}
	consistently with \eqref{generic massless state}-\eqref{massless inner product}.
	
	Thus we want to decompose the massless wavefunctions $\psi_\sigma(\omega,\vec x)$ over a set of wavefunctions $\psi_\sigma(\Delta^*,\vec x)$ carrying elementary representations of $\operatorname{SO}(1,d+1)$ as described in section~\ref{section 2}. This is easily done by making use of the Mellin transform
	\begin{equation}
	\label{Mellin transform}
	\psi_\sigma(\Delta^*,\vec x)=\frac{1}{\sqrt{2\pi}} \int_0^\infty d\omega\, \omega^{\Delta^*-1}\, \psi_\sigma(\omega,\vec x)\,.
	\end{equation}
	Indeed this implements the replacement $-\omega \partial_\omega \leftrightarrow \Delta^*$ mentioned after \eqref{massless rep} such that $\psi_\sigma(\Delta^*,\vec x)$ transforms by an elementary representation of the Lorentz group. While the present discussion is at the level of infinitesimal symmetry transformations, we give the finite version in appendix~\ref{app:1}. For a wavefunction $\psi_\sigma(\omega,\vec x)$ behaving as $O(\omega^{-\alpha})$ in the limit $\omega \to 0$, with $\alpha<\frac{d}{2}$ as imposed by \eqref{zero frequency behavior}, the Mellin transform \eqref{Mellin transform} is holomorphic on the complex half-plane $\text{Re}(\Delta^*)>\alpha$. In particular it is holomorphic in the variable $\Delta^*$ for $\text{Re}(\Delta^*)\geq \frac{d}{2}$. The inverse Mellin transform then provides the sought-for decomposition of the massless wavefunction into elementary representations,
	\begin{equation}
	\psi_\sigma(\omega,\vec x)=\frac{1}{\sqrt{2\pi} i}\int_{c-i\infty}^{c+i\infty} d\Delta^*\, \omega^{-\Delta^*}\, \psi_\sigma(\Delta^*,\vec x)\,, 
	\end{equation}
	with $c$ chosen such that no singularity lies to the right of the integration path. With the appropriate choice $c=d/2$ this can be rewritten
	\begin{equation}
	\label{decomposition massless wavefunction}
	\psi_\sigma(\omega,\vec x)=\frac{1}{\sqrt{2\pi} }\int_{-\infty}^{\infty} d\nu\, \omega^{-\frac{d}{2}+i\nu}\, \psi_\sigma(\nu,\vec x)\,, 
	\end{equation}
	with $\psi_\sigma(\nu,\vec x)$ transforming by the elementary representation $[\Delta^*,\ell]$ with scaling dimension $\Delta^*=\frac{d}{2}-i\nu$ and thus belonging to the continuous principal series. Even though the states $|p(\omega,\vec x)\rangle_\sigma$ and $|\nu,\vec x\rangle_\sigma$ are not normalisable, a similar formal decomposition applies,
	\begin{equation}
	\label{decomposition p}
	|p(\omega,\vec x)\rangle_\sigma=\frac{1}{\sqrt{2\pi}}\int_{-\infty}^{\infty} d\nu\, \omega^{-\frac{d}{2}-i\nu}\, |\nu,\vec x\rangle_\sigma\,, 
	\end{equation}
	with inverse
	\begin{equation}
	\label{inverse p decomposition}
	|\nu,\vec x\rangle_\sigma=\frac{1}{\sqrt{2\pi}}\int_0^\infty d\omega\, \omega^{\frac{d}{2}+i\nu-1} |p(\omega,\vec x)\rangle_\sigma\,.     
	\end{equation}
	Using \eqref{Mellin transform} and \eqref{decomposition p}, the generic massless particle state \eqref{generic massless state} can then be written
	\begin{equation}
	\label{psi decomposition}
	|\psi\rangle = \sum_\sigma \int_{-\infty}^\infty d\nu \int d^d\vec x\, \psi_\sigma(\nu,\vec x) |\nu,\vec x\rangle_\sigma\,,
	\end{equation}
	where we recall that  $|\nu,\vec x\rangle_\sigma$ has scaling dimension $\Delta=\frac{d}{2}+i\nu$ while $\psi_\sigma(\vec x,\nu)$ has conjugate scaling dimension $\Delta^*=\frac{d}{2}-i\nu$. This takes the form of a state \eqref{psi} belonging to the carrier space of a reducible unitary Lorentz representation, with weight multiplicity $\rho(\nu)=1$.
	
	We should check consistency of the Wigner inner product \eqref{massless inner product} with the Lorentz inner product \eqref{Lorentz inner product}. Starting from \eqref{massless inner product} and plugging in the wavefunction decomposition \eqref{decomposition massless wavefunction}, we get
	\begin{equation}
	\label{massless inner product decomposition}
	\begin{split}
	\langle \phi|\psi \rangle&=\frac{1}{2\pi} \sum_\sigma \int d^d\vec x \int_{-\infty}^\infty d\nu d\nu'\, \phi_\sigma(\nu',\vec x)^*\, \psi_\sigma(\nu,\vec x) \int_0^\infty d\omega\, \omega^{i(\nu'-\nu)-1}\\
	&=\sum_\sigma \int_{-\infty}^\infty d\nu \int d^d\vec x \, \phi_\sigma(\nu,\vec x)^*\, \psi_\sigma(\nu,\vec x) \,, 
	\end{split}
	\end{equation}
	where we used the distributional identity
	\begin{equation}
	\int_0^\infty d\omega\, \omega^{i(\nu'-\nu)-1}=2\pi \delta(\nu'-\nu)\,.
	\end{equation}
	The final expression in \eqref{massless inner product decomposition} agrees with the inner product \eqref{Lorentz inner product} that a Hilbert space consisting of the direct integral of Lorentz UIRs with weight $\rho(\nu)=1$ comes equipped with. 
	
	Equation \eqref{decomposition p} provides the decomposition of a massless particle state of definite momentum into states carrying irreducible representations of the Lorentz group. Conversely, it shows how states of the unitary principal series can be assembled such as to form a massless representation of the full Poincar\'e group. From that perspective, unitary translations generated by $U(a)=e^{-i a^\mu \tilde P_\mu}$ connect distinct component Lorentz UIRs.\footnote{Note that $\tilde P_\mu$ itself is unbounded and generically does not produce normalisable states.} We can explicitly work out their action using \eqref{inverse p decomposition},  
	\begin{equation}
	\label{action translation}
	\begin{split}
	U(a) |\nu,\vec x\rangle_\sigma&=\frac{1}{\sqrt{2\pi}}\int_0^\infty d\omega\, \omega^{\frac{d}{2}+i\nu-1} e^{-i a^\mu \tilde P_\mu} |p(\omega,\vec x)\rangle_\sigma\\
	&=\frac{1}{\sqrt{2\pi}}\int_0^\infty d\omega\, \omega^{\frac{d}{2}+i\nu-1} e^{-i\omega a^\mu q_\mu(\vec x)} |p(\omega,\vec x)\rangle_\sigma\\
	&=\frac{1}{(2\pi)^{3/2}} \int_{-\infty}^\infty d\nu'\, \frac{\Gamma(\epsilon+i(\nu-\nu'))} {(ia^\mu q_\mu(\vec x)+\epsilon)^{i(\nu-\nu')}} \int_0^\infty d\omega\, \omega^{\frac{d}{2}+i\nu'-1} |p(\omega,\vec x)\rangle_\sigma\\
	&=\frac{1}{2\pi} \int_{-\infty}^\infty d\nu'\, \frac{\Gamma(\epsilon+i(\nu-\nu'))} {(ia^\mu q_\mu(\vec x)+\epsilon)^{i(\nu-\nu')}}\, |\nu',\vec x\rangle_\sigma\,\\
	&=\frac{1}{2\pi} \int_{-\infty}^\infty d\nu'\, {\rm e}^{-\frac{\pi}{2}(\nu-\nu')}\frac{\Gamma(\epsilon+i(\nu-\nu'))} {(-a^\mu q_\mu(\vec x)+i\epsilon)^{i(\nu-\nu')}}\, |\nu',\vec x\rangle_\sigma\,,
	\end{split} 
	\end{equation}
	where we used the Mellin representation of the exponential function
	\begin{equation}
	e^{-i\omega x-\epsilon}=\frac{1}{2\pi}\int_{-\infty}^\infty d\nu\, \Gamma(\epsilon-i\nu) (i\omega x+\epsilon)^{i\nu}\,, \qquad \omega x\in\mathbb{R},\quad \epsilon>0\,.
	\end{equation}
	
	Hence we see that translations map a state of a given scaling dimension to states of all scaling dimensions, though still restricted to the unitary principal series. Of course this had to be the case, otherwise the statement that the principal series provides a decomposition of the full massless Poincar\'e representation could not have been correct. We should contrast this result with the widespread statement that translations take states out of the principal series, which follows from unwarranted application of the unbounded operator $\tilde P_\mu$. Indeed one naively finds, for instance,
	\begin{equation}
	\tilde P^0 |\Delta,\vec 0\,\rangle=\frac{1}{\sqrt{2\pi}}\int_0^\infty d\omega\, \omega^{\Delta} |p(\omega,\vec x)\rangle=|\Delta+1,\vec 0\,\rangle\,,
	\end{equation}
	such that if one starts with $\Delta \in \frac{d}{2}+i\mathbb{R}$, then $\Delta +1$ indeed lies off the unitary principal series. Generation of such non-normalisable states is avoided by working with the unitary operators $U(a)=e^{-ia^\mu \tilde P_\mu}$. 
	
	\section{Decomposition of massive particles}
	\label{section 4}
	Now we turn to decomposition of the unitary irreducible representations of the Poincar\'e group $\operatorname{ISO}(1,d+1)$ that describe massive particle states. Let us start by recalling Wigner's construction of the massive UIRs of the Poincaré
	group~\cite{Wigner:1939cj}. Within an irreducible representation, the quadratic Casimir operator $\mathcal{C}_2^{\mathfrak{iso}(1,d+1)}=\tilde P^\mu \tilde P_\mu$ should be proportional to the identity, and we thus require
	\begin{equation}
	\tilde P^\mu \tilde P_\mu=-m^2\,.
	\end{equation}
	Henceforth one starts with a reference momentum $k^\mu$ satisfying the above constraint, which can be chosen to be the momentum in its rest frame for simplicity,
	\begin{equation}
	\label{kmu}
	k^\mu= m\,\delta^\mu_0\,.
	\end{equation}
	The little group leaving this reference momentum invariant is spanned by spatial rotations and translations that are generated by $\langle J_{ab}, \tilde P_\mu \rangle$, respectively, with $a,b=1,...,d+1$ the purely spatial indices.
	The representation of the little group, which will be used to induce the representation of the full Poincar\'e group, is given by
	\begin{equation}
	J_{ab}\, |k\rangle_\sigma=(\Sigma_{ab})\indices{_\sigma^{\sigma'}} |k\rangle_{\sigma'}\,, \qquad \tilde P_\mu\, |k\rangle_\sigma=k_\mu\, |k\rangle_\sigma\,,
	\end{equation}
	where $(\Sigma_{ab})\indices{_\sigma^{\sigma'}}$ is a set of hermitian matrices furnishing a unitary irreducible representation of $\operatorname{SO}(d+1)$. The induced representation is then obtained by boosting $|k \rangle_\sigma$ from its rest frame to a generic momentum frame. More specifically, given the Lorentz transformation $L(\hat p)\indices{^\mu_\nu}$ which boosts $k^\mu$ into the generic momentum $p^\mu=m \hat p^\mu$ with $\hat p^2=-1$,
	\begin{equation}
	p^\mu=L(\hat p)\indices{^\mu_\nu}\, k^\nu\,,
	\end{equation}
	with components and inverse components explicitly given by
	\begin{equation}
	\label{boost matrix}
	\begin{split}
	L(\hat p)\indices{^0_0}&=\hat p^0\,,\\
	L(\hat p)\indices{^a_0}&=L(\hat p)\indices{^0_a}=\hat p^a\,,\\
	L(\hat p)\indices{^a_b}&=\delta^a_b+(\hat p^0+1)^{-1}\, \hat p^a \hat p_b\,,
	\end{split}
	\end{equation}
	and 
	\begin{equation}
	\begin{split}
	L^{-1}(\hat p)\indices{^0_0}&=-\hat p_0\,,\\
	L^{-1}(\hat p)\indices{^a_0}&=L^{-1}(\hat p)\indices{^0_a}=-\hat p_a\,,\\
	L^{-1}(\hat p)\indices{^a_b}&=\delta^a_b+(\hat p^0+1)^{-1}\, \hat p^a \hat p_b\,,
	\end{split}
	\end{equation} 
	respectively, one defines the boosted state by
	\begin{equation}
	|p\rangle_\sigma\equiv U(L(\hat p))\, |k\rangle_\sigma\,.
	\end{equation}
	Here $U(L(\hat p))$ denotes a unitary representation of the boost element $L(\hat p)$, to be determined following the method of induced representations. For a generic Lorentz transformation of the basis states $|p\rangle_\sigma$, the latter yields \cite{Weinberg:1995mt} 
	\begin{equation}
	\label{abstract Wigner}
	U(\Lambda)|p\rangle_\sigma=D(W(\Lambda,\hat p)^{-1})\indices{_\sigma^{\sigma'}} |\Lambda p\rangle_{\sigma'}\,,
	\end{equation}
	where $W(\Lambda,\hat p) \in \operatorname{SO}(d+1)$ is a rotation leaving the reference momentum $k^\mu$ invariant, given by
	\begin{equation}
	\label{Wigner rotation abstract}
	W(\Lambda,\hat p)=L^{-1}(\Lambda \hat p) \Lambda L(\hat p)\,,
	\end{equation}
	and where $D(W(\Lambda,\hat p))$ is its unitary spin matrix representation. 
	
	In the Hilbert space carrying this representation, a generic state takes the form
	\begin{equation}
	\label{generic massive state}
	|\psi \rangle=\sum_\sigma \int [d^{d+1} \hat p]\, \psi_\sigma(\hat p)\, |p\rangle_\sigma \,,
	\end{equation}
	with $\psi_\sigma(\hat p)$ any complex wavefunction normalisable with respect to the inner product
	\begin{equation}
	\label{massive inner product}
	\langle \phi | \psi \rangle=\sum_\sigma \int [d^{d+1} \hat p]\, \phi_\sigma(\hat p)^*\, \psi_\sigma(\hat p)\,,    
	\end{equation}
	and with $[d^{d+1} \hat p]$ the Lorentz-invariant measure on the hyperboloid $\mathbb{H}^{d+1}$,
	\begin{equation}
	[d^{d+1} \hat p]=\frac{d^{d+1}\hat{\mathbf{p}}}{\sqrt{1+||\hat{\mathbf{p}}||^2}}\,, \qquad \hat p^\mu=(\sqrt{1+||\hat{\mathbf{p}}||^2}\,, \hat{\mathbf{p}})\,.
	\end{equation}
	The complex wavefunctions $\psi_\sigma(\hat p)$ again naturally carry the conjugate unitary representation, isomorphic to original representation. Thus the Hilbert space is isomorphic to the space of square integrable complex wavefunctions $L^2(\mathbb{H}^{d+1},[d^{d+1} \hat p])$. Even though the states $|p\rangle_\sigma$ do not strictly belong to this Hilbert space, one can assign to them the distributional inner product 
	\begin{equation}
	{}_{\sigma'}\langle p_1|p_2\rangle_\sigma=\sqrt{1+||\hat{\mathbf{p}}_1||^2}\, \delta(\hat{\mathbf{p}}_1-\hat{\mathbf{p}}_2)\, \delta_{\sigma\sigma'}\equiv \delta(\hat p_1-\hat p_2)\,,
	\end{equation}
	consistently with \eqref{generic massive state}-\eqref{massive inner product}.
	
	We would like to decompose the wavefunctions $\psi_\sigma(\hat p)$ into components carrying irreducible representations of the Lorentz group $\operatorname{SO}(1,d+1)$ spanned by $\langle J_{\mu\nu} \rangle$, which precisely generate the isometries of the momentum mass-shell $\HH^{d+1}$. This amounts to a problem of harmonic decomposition of $L^2(\mathbb{H}^{d+1},[d^{d+1} \hat p])$, wherein the momentum mass-shell is viewed as the homogeneous space
	\begin{equation}
	\mathbb{H}^{d+1} \cong \frac{\operatorname{SO}(1,d+1)}{\operatorname{SO}(d+1)}\,.  
	\end{equation}
	The analysis we present here will be rather practical, and we refer the interested reader to \cite{Barut:1986dd,Dobrev:1977qv} for a detailed discussion of harmonic analysis in relation to group theory.  
	We first note that the set of normalisable wavefunctions $\psi_\sigma(\hat p)$ does not provide an irreducible carrier space since the action of the quadratic Casimir $\mathcal{C}_2^{\mathfrak{so}(1,d+1)}=\frac{1}{2}J_{\mu\nu}J^{\mu\nu}$ on these wavefunctions is not proportional to the identity, but rather to the  Laplace--Beltrami operator $\nabla^2$ on $\HH^{d+1}$, 
	\begin{equation}
	\mathcal{C}_2^{\mathfrak{so}(1,d+1)} \psi_\sigma(\hat p)=\left( \nabla^2+\mathcal{C}_2^{\mathfrak{so}(d+1)} \right) \psi_\sigma(\hat p)=\left( \nabla^2+s(s+d-1) \right) \psi_\sigma(\hat p)\,,
	\end{equation}
	where in the second equality we restricted our attention to totally symmetric spin-$s$ representations of $\operatorname{SO}(d+1)$. 
	Hence the idea is to decompose $\psi_\sigma(\hat p)$ over a set of wavefunctions $G^{(s)}_{\nu,\ell}(\hat p;...)$ with definite Casimir value \eqref{Casimir principal series} and carrying an irreducible Lorentz representation of the principal series, i.e., satisfying the Laplace--Beltrami equation
	\begin{equation}
	\label{Laplace equation}
	\left(\nabla^2+s(s+d-1)\right) G^{(s)}_{\nu,\ell}(\hat p;...)=\left(-(\nu^2+d^2/4)+\ell(\ell+d-2) \right) G^{(s)}_{\nu,\ell}(\hat p;...)\,.
	\end{equation}
	Here the dots $...$ refer to additional labels for the complete set of such functions $G^{(s)}_{\nu,\ell}(\hat p;...)$, which will be given momentarily. These labels will be analogues of the quantum numbers $(j,m)$ labeling spherical harmonics $Y_{jm}(\Omega)$ that provide a basis of scalar functions on the sphere. Here we decompose wavefunctions over $\HH^{d+1}$, whose noncompact nature requires the use of continuous quantum numbers rather than discrete ones.
	
	The first point to discuss is the appearance of two distinct spin parameters $s$ and $\ell$, the first one being associated to the little group $\operatorname{SO}(d+1)$ of the massive particle, and the second one being associated with the subgroup $\operatorname{M}=\operatorname{SO}(d)$ defined in \eqref{Bruhat} and characterising the elementary representations of the Lorentz group. The decomposition of the particle representation into Lorentz UIRs will thus feature a decomposition of the tensor irrep $V^{\operatorname{SO}(d+1)}_s$ into the tensor irreps $V^{\operatorname{SO}(d)}_\ell$, simply given by
	\begin{equation}
	\label{spin decomposition}
	V^{\operatorname{SO}(d+1)}_s=\bigoplus_{\ell=0}^s V^{\operatorname{SO}(d)}_\ell\,,
	\end{equation}
	with multiplicity one. For ease of exposition, we will first describe the spinless case $s=\ell=0$ before treating generic integer spin $s$, which requires introducing some heavier technology. We will then analyse the half-integer spin case, which requires to consider spinor representations of $\operatorname{SO}(d+1)$ and $\operatorname{SO}(d)$. 
	
	\subsection{Zero spin}
	For $s=\ell=0$, the set of regular solutions to the Laplace--Beltrami equation \eqref{Laplace equation}  is given by
	\begin{equation}
	\label{bulk-boundary propagators}
	G_{\nu}(\hat p;\vec x)=\frac{1}{(-\hat p \cdot q(\vec x))^{\Delta_\nu}}\,, \qquad \Delta_\nu=\frac{d}{2}+i\nu\,,
	\end{equation}
	with $\vec x \in \mathbb{R}^d$ a continuous label and $q^\mu(\vec x)$ the null vector given in \eqref{massless momentum parametrisation}. 
	The solutions \eqref{bulk-boundary propagators} with $\nu \in \mathbb{R}^+$ and $\vec x \in \mathbb{R}^d$ form a basis of normalisable wavefunctions in $L^2(\mathbb{H}^{d+1},[d^{d+1} \hat p])$. Within this functional space, a resolution of the identity is given by \cite{Costa:2014kfa}
	\begin{equation}
	\int_0^\infty d\nu\, \mu(\nu) \int d^d\vec x\, G_{-\nu}(\hat p_1;\vec x)\, G_{\nu}(\hat p_2;\vec x)=\delta(\hat p_1-\hat p_2)\,,
	\label{complScalar}
	\end{equation}
	with $\mu(\nu)$ the Plancherel measure
	\begin{equation}
	\label{Plancherel scalar}
	\mu(\nu)=\frac{\Gamma(\frac{d}{2}+i\nu)\Gamma(\frac{d}{2}-i\nu)}{4\pi^{d+1}\Gamma(i\nu)\Gamma(-i\nu)}\,.
	\end{equation}
	This allows to decompose a generic wavefunction $\psi(\hat p)$ as
	\begin{equation}
	\label{wavefunction decomposition}
	\psi(\hat p)=\int_0^\infty d\nu\, \mu(\nu) \int d^d\vec x\, G_{\nu}(\hat p;\vec x)\, \psi(\nu,\vec x)\,,
	\end{equation}
	with
	\begin{equation}
	\label{wavefunction decomposition bis}
	\psi(\nu,\vec x)\equiv \int [d^{d+1}\hat p]\, G_{-\nu}(\hat p;\vec x)\, \psi(\hat p)\,.
	\end{equation}
	Even though they are not normalisable, a similar decomposition can be written in terms of basis states $|p\rangle$ and $|\nu,\vec x\rangle$, namely
	\begin{equation}
	\label{state decomposition}
	|p\rangle=\int_0^\infty d\nu\, \mu(\nu) \int d^d\vec x\, G_{-\nu}(\hat p;\vec x)\, |\nu,\vec x\rangle\,,
	\end{equation}
	with inverse
	\begin{equation}
	\label{nu,x}
	|\nu,\vec x \rangle= \int [d^{d+1}\hat p]\, G_{\nu}(\hat p;\vec x)\, |p\rangle\,.
	\end{equation}
	Let us now explicitly show that the states \eqref{nu,x} do indeed transform according to the Lorentz representations of the principal series described in section~\ref{section 2}. Acting with a Lorentz transformation $U(\Lambda)$, we have
	\begin{equation}
	\label{U(Lambda) nu}
	\begin{split}
	U(\Lambda)|\nu,\vec x \rangle=\int [d^{d+1}\hat p]\, G_{\nu}(\hat p;\vec x)\, |\Lambda p\rangle=\int [d^{d+1}\hat p]\, G_{\nu}(\Lambda^{-1} \hat p;\vec x)\, |p\rangle\,,
	\end{split}
	\end{equation}
	where we used the transformation of the momentum state $U(\Lambda)|p\rangle=|\Lambda p\rangle$, and the Lorentz invariance of the integral measure. At this point we recall that the Lorentz group $\operatorname{SO}(1,d+1)$ is naturally realised as the group of conformal isometries of $\mathbb{R}^d$, which we abstractly denote $\vec x \mapsto \vec x'=\Lambda \vec x$ for any $\vec x \in \mathbb{R}^d$. Under this transformation the null vector \eqref{massless momentum parametrisation} satisfies \cite{Pasterski:2017kqt} 
	\begin{equation}
	\label{relation}
	q^\mu(\vec{x}\,')=\Omega(x) \Lambda^\mu{}_\nu q^\nu(\vec{x})\,, \qquad  \Omega(x)=\left|\frac{\,\partial{\vec{x}\,'}}{\partial\vec{x}}\right|^{1/d}\,,
	\end{equation}
	which allows us to write
	\begin{equation}
	G_{\nu}(\Lambda^{-1} \hat p;\vec x)=\frac{1}{\left(-\hat p \cdot \Lambda q(\vec x) \right)^{\Delta_\nu}}=\left(\frac{\Omega(x)}{-\hat p \cdot q(\vec x')}\right)^{\Delta_\nu} \,.
	\end{equation}
	Plugging this back into \eqref{U(Lambda) nu} yields
	\begin{equation}
	\begin{split}
	U(\Lambda)|\nu,\vec x \rangle=\Omega(x)^{\Delta_\nu} |\nu,\vec x\,'\rangle\,,
	\end{split}
	\end{equation}
	which is the transformation law \eqref{Lorentz finite transfo} associated with a scalar representation of the principal series.
	
	Using \eqref{wavefunction decomposition}-\eqref{nu,x}, the generic one-particle state \eqref{generic massive state} can be written
	\begin{equation}
	\label{generic psi decomposition}
	|\psi\rangle=\int_0^\infty d\nu\, \mu(\nu) \int d^d\vec x\,\psi(\nu,\vec x)|\nu,\vec x\rangle\,,
	\end{equation}
	which is of the general form \eqref{psi} with $\rho(\nu)=\Theta(\nu)\mu(\nu)$. As for massless particles, one can check that the Lorentz inner product \eqref{Lorentz inner product} is recovered from the particle inner product \eqref{massive inner product} together with the decomposition \eqref{generic psi decomposition}. Indeed we can write
	\begin{align}
	\langle \phi|\psi\rangle=\int [d^{d+1}\hat p]\, \phi(\hat p)^* \psi(\hat p)=\int_0^\infty d\nu\, \mu(\nu) \int d^d\vec x\, \psi(\nu,\vec x)^*\, \psi(\nu,\vec x)\,,
	\end{align}
	using the completeness relation \eqref{complScalar}, or the orthogonality relation
	\begin{equation}
	\int [d^{d+1}\hat p]\, G_{-\nu'}(\hat p;\vec w) G_{\nu}(\hat p;\vec x)=\frac{\delta(\nu-\nu')\delta(\vec w-\vec x)}{\mu(\nu)}\,, \qquad \nu,\nu' \in \mathbb{R}^+\,.
	\end{equation}
	
	Finally, we wish to compute the action of translations $U(a)=e^{-ia^\mu \tilde P_\mu}$ on the basis states $|\nu,\vec x\rangle$ such that the latter indeed assemble into the carrier space of the massive particle representation. This can be done using \eqref{state decomposition}-\eqref{nu,x}, yielding
	\begin{equation}
	\label{translations massive}
	\begin{split}
	U(a) |\nu,\vec x\rangle&= \int [d^{d+1}\hat p]\, G_{\nu}(\hat p;\vec x)\, e^{-im a^\mu \hat p_\mu} |p\rangle\\
	&=\int_0^\infty d\nu'\, \mu(\nu') \int d^d\vec y\,   A_3(a^\mu;\nu,\vec x;\nu',\vec y)\, |\nu',\vec y\rangle\,,
	\end{split}
	\end{equation}
	with 
	\begin{equation}
	\label{A3}
	A_3(a^\mu;\nu,\vec x;\nu',\vec y)\equiv \int [d^{d+1}\hat p]\, G_\nu(\hat p;\vec x) G_{-\nu'}(\hat p;\vec y)\, e^{-ima^\mu\hat p_\mu}\,.
	\end{equation}
	Like in the massless case, equation \eqref{translations massive} shows that translations mix states of all scaling dimensions within the principal series. The coefficient function \eqref{A3} is recognised as a `mixed' massive celestial three-point function, with one leg $a^\mu \in \mathbb{M}^{d+2}$ still in `position space'. In appendix~\ref{appendix B} we explain how to explicitly evaluate this three-point function. 
	
	\subsection{Index-free and ambient space formalism}
	The treatment of symmetric tensor representation $V_s^{\operatorname{SO}(d+1)}$, and their decomposition \eqref{spin decomposition} in terms of representations $V_\ell^{\operatorname{SO}(d)}$, will be facilitated by the use of \textit{index-free} and \textit{ambient space} formalisms. These were heavily used by the authors of \cite{Costa:2011mg,Costa:2014kfa} for their description of spinning representations in the context of the AdS/CFT correspondence, who incidentally also provided formulae which will play a key role in the present analysis. 
	
	Up to this point we have not given an explicit parametrisation of the representations $V_\ell^{\operatorname{SO}(d)}$ appearing in the above discussion. For later use it is useful to introduce their polynomial realisation which is the basis of the \textit{index-free} formalism. While we restrict here to totally symmetric traceless spin-$\ell$ tensor representations, a more general discussion can be found for example in \cite{Bekaert:2006py}. Given the totally symmetric and traceless (with respect to $\delta_{ij}$) tensor field $f_\sigma(\vec x)=f_{i_1...i_\ell}(\vec x)$ in $\mathbb{R}^d$, we can construct a homogeneous polynomial of degree $\ell$ by contracting it with an auxiliary `polarisation' vector $\vec z \in \mathbb{R}^d$,
	\begin{equation}
	\label{polynomial section 2}
	f(\vec x,\vec z)\equiv f_{i_1...\,i_\ell}(\vec x)\, z^{i_1}...\,z^{i_\ell}\,,
	\end{equation}
	which can be shown to satisfy
	\begin{equation}
	\label{psi(z) conditions}
	\left( \vec z \cdot \partial_{\vec z}-\ell\right) f(\vec x,\vec z)=0\,, \qquad \left(\partial_{\vec z} \cdot \partial_{\vec z} \right)f(\vec x,\vec z)=0\,.
	\end{equation}
	Conversely, a polynomial satisfying \eqref{psi(z) conditions} is equivalent to a totally symmetric and traceless tensor of rank $\ell$. Resctricted to the unit sphere $S^{d-1} \subset \mathbb{R}^d$ where $\vec z \cdot \vec z=1$, such polynomials coincide with spherical harmonics of degree $\ell$. The action of $\operatorname{SO}(d)$ on $f(\vec x,\vec z)$ is now explicitly implemented as a rotation of the polarisation vector $\vec z$, which infinitesimally reads
	\begin{equation}
	\Sigma_{ij}\, f(\vec x,\vec z)=-i\left(z_i \frac{\partial}{\partial z^j}-z_j \frac{\partial}{\partial z^i}\right) f(\vec x, \vec z)\,.
	\end{equation}
	Following \cite{Costa:2011mg} it is convenient to complexify the polarisation vector $\vec z$ and consider the restriction of $f(\vec x,\vec z)$ to the submanifold $\vec z \cdot \vec z=0$. This effectively removes any potential trace term such that the second condition in \eqref{psi(z) conditions} becomes unnecessary. 
	
	The tensor field $f_{i_1...\,i_\ell}(\vec x)$ in $\mathbb{R}^d$ can also be encoded by a tensor field $f_{\mu_1...\,\mu_\ell}(q)$ defined on the null cone $\operatorname{LC}^+_{d+1} \subset \mathbb{M}^{d+2}$ located at $q^2\equiv \eta_{\mu\nu} q^\mu q^\nu=0$, which is totally symmetric, traceless with respect to $\eta_{\mu\nu}$, and transverse with respect to $q^\mu$, through the pull-back
	\begin{equation}
	\label{pullback}
	f_{i_1...\,i_\ell}(\vec x)=\frac{\partial q^{\mu_1}}{\partial x^{i_1}}\,...\, \frac{\partial q^{\mu_\ell}}{\partial x^{i_\ell}} f_{\mu_1...\,\mu_\ell}(q)\,,
	\end{equation}
	together with the parametrisation $q^\mu(\vec x)$ given in \eqref{massless momentum parametrisation}. This constitutes the basis of the \textit{ambient space} formalism. Note that the extension of $f_{\mu_1...\,\mu_\ell}(q)$ off the null cone $q^2=0$ is inconsequential. In order to get rid of the indices, we again construct a homogeneous polynomial of degree $\ell$ by contracting with the complex auxiliary vector $Z^\mu$,
	\begin{equation}
	\label{f(q,Z) definition}
	f(q,Z)\equiv f_{\mu_1...\,\mu_\ell}(q)\, Z^{\mu_1}\,...\, Z^{\mu_\ell}\,.
	\end{equation}
	The polynomial $f(\vec x,\vec z)$ is then obtained via
	\begin{equation}
	f(\vec x,\vec z)=f(q,Z)\,,
	\end{equation}
	with $q^\mu$ given as in \eqref{massless momentum parametrisation} and
	\begin{equation}
	\label{Z vector}
	Z^\mu=(\vec x\cdot \vec z,\vec z,-\vec x\cdot \vec z)\,.
	\end{equation}
	The polarisation vector $Z^\mu$ satisfies $Z^2=\vec z \cdot \vec z$ and $Z \cdot q=0$. 
	Conversely, it was shown in \cite{Costa:2011mg} that a homogeneous polynomial $f(q,Z)$ of degree $\ell$ in $\mathbb{M}^{d+2}$, when restricted to the submanifold $Z^2=Z \cdot q=q^2=0$, unambiguously encodes a symmetric traceless tensor field $f_{i_1...\,i_\ell}(\vec x)$ in $\mathbb{R}^d$. Its explicit construction proceeds through \eqref{pullback} and
	\begin{equation}
	\label{components f}
	f_{\mu_1...\,\mu_\ell}(q)=\frac{1}{\ell!(h-1)_\ell} (D_Z)_{\mu_1}...\,(D_Z)_{\mu_\ell}\, f(q,Z)\,,
	\end{equation}
	with $h\equiv d/2$, $(a)_\ell\equiv \Gamma(a+\ell)/\Gamma(a)$ the Pochhammer symbol, and
	\begin{equation}
	(D_Z)_\mu\equiv \left(h-1+Z \cdot \frac{\partial}{\partial Z}\right)\frac{\partial}{\partial Z^\mu}\,.
	\end{equation}
	We note that this derivative operator can be used to implement tensor contractions at the level of the encoding polynomials, such as
	\begin{equation}
	f_{i_1...i_\ell}(\vec x) g^{i_1...i_\ell}(\vec x)=\frac{1}{\ell!(h-1)_\ell} f(q,D_Z) g(q,Z)\,.
	\end{equation}
	
	A similar description exists for the totally symmetric traceless spin-$s$ tensor representations of $\operatorname{SO}(d+1)$ \cite{Costa:2014kfa}. Given the symmetric and traceless tensor field $h_\sigma(\hat p)=h_{a_1\,...a_s}(\hat p)$ in $\mathbb{H}^{d+1}$,
	there is a symmetric traceless tensor field $H_{\mu_1...\,\mu_s}(\hat p)$ in $\mathbb{M}^{d+2}$ such that\footnote{ Note that the formulae in \cite{Costa:2014kfa} are written with coordinate indices so that $\partial \hat p^{\mu}/\partial x^{a}$ appears in place of $E^\mu_a$, while \eqref{h from H} is written with orthonormal frame indices.}
	\begin{equation}
	\label{h from H}
	h_{a_1...\,a_s}(\x)=E(\hat p)^{\mu_1}_{a_1}\,...\, E(\hat p)^{\mu_s}_{a_s} H_{\mu_1...\,\mu_s}(\hat p)\,,
	\end{equation}
	with $E(\hat p)^\mu_a$ an orthonormal frame field satisfying $E^\mu_a\, \eta_{\mu\nu} E^\nu_b=\delta_{ab}$ and $E(\hat p)^\mu_a\, \hat p_\mu=0$. The latter can be explicitly chosen as
	\begin{equation}
	\label{frame E}
	E(\hat p)^0_a=\hat p_a\,, \qquad E(\hat p)^b_a=\delta^b_a+(\hat p^0+1)^{-1}\, \hat p^b \hat p_a\,,
	\end{equation}
	with inverse $E^{-1}(\hat p)^a_\mu$ given by
	\begin{equation}
	\label{inverse frame}
	E^{-1}(\hat p)^a_0=-\hat p^a\,, \qquad E^{-1}(\hat p)^a_b=\delta^a_b+(\hat p^0+1)^{-1}\, \hat p^a \hat p_b\,.
	\end{equation}
	They provide a resolution of the identity through the formula
	\begin{equation}
	\label{resolution of identity E}
	\delta^\mu_\nu=-\hat p^\mu \hat p_\nu+E(\hat p)^\mu_a\, E^{-1}(\hat p)^a_\nu\,. 
	\end{equation}
	
	Again we construct a homogeneous polynomial of degree $s$ by contracting with the complex polarisation vector $W^\mu$,
	\begin{equation}
	\label{H polynomial}
	H(\hat p,W)\equiv H_{\mu_1...\,\mu_s}(\hat p)\, W^{\mu_1}\,...\, W^{\mu_s}\,,
	\end{equation}
	satisfying
	\begin{equation}
	\left(W \cdot \partial_W-s\right)H(\hat p,W)=0\,, \qquad \partial_W\cdot \partial_W\, H(\hat p,W)=0\,.
	\end{equation}
	The action of $\operatorname{SO}(d+1)$ on $H(\hat p,W)$ is now explicitly implemented as rotations of the polarisation vector $W^\mu$, and infinitesimally reads 
	\begin{equation}
	\Sigma_{ab}\, H(\hat p,W)=-i E^\mu_a E^\nu_b\left(W_\mu \frac{\partial}{\partial W^\nu}-W_\nu \frac{\partial}{\partial W^\mu}\right) H(\hat p, W)\,.
	\end{equation}
	Conversely, a homogeneous polynomial $H(\hat p,W)$ of degree $s$ in $\mathbb{M}^{d+2}$, when restricted to the submanifold $W^2=W \cdot \hat p=\hat p^2+1=0$, unambiguously encodes a symmetric traceless tensor field $h_{a_1...\,a_s}(\hat p)$ in $\mathbb{H}^{d+1}$. Its explicit construction proceeds through \eqref{h from H} and 
	\begin{equation}
	H_{\mu_1...\,\mu_s}(\hat p)=\frac{1}{s!\left(\frac{d-1}{2}\right)_s} (K_W)_{\mu_1}\,...\, (K_W)_{\mu_s}\, H(\hat p,W)\,,
	\end{equation}
	with 
	\begin{equation}
	\label{K mu}
	(K_W)_\mu\equiv \left(\frac{d-1}{2}+W \cdot \frac{\partial}{\partial W} \right)\frac{\partial}{\partial W^\mu}\,.
	\end{equation}
	Finally, the covariant derivative acting tangentially to $\HH^{d+1} \subset \mathbb{M}^{d+2}$ takes the form
	\begin{equation}
	\nabla_\mu=\frac{\partial}{\partial \hat p^\mu}+\hat p_\mu \left(\hat p \cdot \frac{\partial}{\partial \hat p}\right)+W_\mu\left(\hat p \cdot \frac{\partial}{\partial W} \right)\,.
	\end{equation}
	The divergence of the tensor $H_{\mu_1...\mu_s}(\hat p)$, in index-free notation, can then be computed as
	\begin{equation}
	(\nabla \cdot H)(\hat p,W)=\frac{1}{s\left(\frac{d-3}{2}+s\right)} \nabla \cdot K\, H(\hat p,W)\,.
	\end{equation}
	Armed with this technology we are ready to resume our study of massive spinning particles and their decomposition into Lorentz UIRs.
	
	\subsection{Integer spin}
	Adopting the ambient space formalism with index-free notation, the generalisation of the completeness relation \eqref{complScalar} to the space of spin-$s$ normalisable wavefunctions has been given in \cite{Costa:2014kfa}, 
	\begin{equation}
	\label{Costa et al}
	\begin{split}
	\sum_{\ell=0}^s \frac{1}{\ell!(h-1)_\ell} \int_{0}^{\infty}d\nu\, \mu_{s,\ell}(\nu) \int d^d \vec x\, G^{(s)}_{\nu,\ell}(\p_1, W_1; q,D_Z) G_{-\nu,\ell}^{(s)}(\p_2, W_2; q,Z)\\
	=(W_1\cdot W_2)^s\, \delta(\p_1, \p_2)\,,
	\end{split}
	\end{equation}
	where the null vector $q^\mu=q^\mu(\vec x)$ implicitly depends on the integration variable $\vec x \in \mathbb{R}^d$ through \eqref{massless momentum parametrisation}, and 
	\begin{equation}
	\begin{split}
	\mu_{s,\ell}(\nu)=\frac{2^{s-\ell}(\ell+1)_{s-\ell} \left(\ell+\frac{d-1}{2} \right)_{s-\ell} }{(s-\ell)!  (2\ell +d-1)_{s-\ell} \left(\ell+h-i \nu\right)_{s-\ell} \left(\ell +h+i \nu\right)_{s-\ell}}\\
	\times \left|\frac{\ell+h+i\nu-1}{h+i\nu-1}\right|^2 \mu(\nu)\,,
	\end{split}
	\end{equation}
	with $\mu(\nu)$ as given in \eqref{Plancherel scalar}.
	The kernel $G^{(s)}_{\Delta,\ell}(\p, W; q,Z)$ is a homogeneous polynomial of degree $s$ in $W$ and degree $\ell$ in $Z$, and thus encodes a bi-tensor $G^{(s)}_{\Delta,\ell}(\p; q)_{a_1...\,a_s;\,i_1...\,i_\ell}$ traceless and symmetric in both sets of indices. Its explicit expression is
	\begin{equation}
	\label{spinning intertwiner}
	G^{(s)}_{\nu,\ell}(\p, W; q,Z)=(W \cdot \nabla)^{s-\ell}\, \Pi_{\nu,\ell}(\hat p,W;q,Z)\,,
	\end{equation}
	given in terms of the spin-$\ell$ `boundary-bulk propagator'
	\begin{equation}
	\Pi_{\nu, \ell}(\hat{p}, W; q, Z)\equiv \frac{\left(- (\hat{p}\cdot q)(W\cdot Z)+(\hat p\cdot Z)( q\cdot W)\right)^\ell}{\left(-\, \hat{p}\cdot q\right)^{\Delta_\nu+\ell}}\,, \qquad \Delta_\nu=h+i\nu\,.
	\label{btb form}
	\end{equation}
	With respect to the ambient coordinates $\hat p^\mu$, the latter satisfies a Laplace--Beltrami equation on $\HH^{d+1}$ as well as a divergence-free condition, 
	\begin{align}
	\label{Laplace Pi}
	\left[\nabla^2-\Delta_\nu(\Delta_\nu-d)+\ell\right]\Pi_{\nu, \ell}(\hat{p}, W; q, Z)=0\,,\\
	\nabla \cdot K\, \Pi_{\nu, \ell}(\hat{p}, W; q, Z)=0\,. 
	\end{align}
	Using \eqref{Laplace Pi} and the commutation relation\cite{Costa:2014kfa}
	\begin{equation}
	\left[\nabla^2,(W \cdot \nabla)^n \right]=-n\left(d-1+2 W\cdot \partial_W-n\right) (W\cdot \nabla)^n\,,
	\end{equation}
	we can easily show
	\begin{equation}
	\begin{split}
	\nabla^2 G^{(s)}_{\nu,\ell}(\p, W; q,Z)&=\left[\Delta_\nu(\Delta_\nu-d)-\ell-(s-\ell)(d-1+s+\ell) \right]G^{(s)}_{\nu,\ell}(\p, W; q,Z)\\
	&=\left[\Delta_\nu(\Delta_\nu-d)-s(s+d-1)+\ell(\ell+d-2)\right]G^{(s)}_{\nu,\ell}(\p, W; q,Z)\,.
	\end{split}
	\end{equation}
	We observe that this is exactly the equation \eqref{Laplace equation} that wavefunctions carrying a Lorentz UIR $[\Delta_\nu,\ell]$ of the spinning principal series must satisfy. It should now become clear that the summation $\sum_{\ell=0}^s$ in \eqref{Costa et al} implements the decomposition \eqref{spin decomposition}. 
	
	As in the spinless case, the completeness relation \eqref{Costa et al} allows us to decompose a generic spin-$s$ wavefunction $\psi(\hat p,W)$ as
	\begin{equation}
	\label{psi(p,W) decomposition}
	\psi(\hat p,W)
	=\sum_{\ell=0}^s \frac{1}{\ell!(h-1)_\ell} \int_{0}^{\infty}d\nu\, \mu_{s,\ell}(\nu) \int d^d \vec x\, G^{(s)}_{\nu,\ell}(\p,W; q,D_Z)\, \psi^{(\ell)}(\nu,q,Z)\,,
	\end{equation}
	with
	\begin{equation}
	\label{psi(nu,q,Z) decomposition}
	\psi^{(\ell)}(\nu,q,Z)=\frac{1}{s!\left(\frac{d-1}{2} \right)_s} \int [d^{d+1}\hat p]\, G_{-\nu,\ell}^{(s)}(\p,K_W; q,Z)\, \psi(\hat p,W)\,.
	\end{equation}
	While $\psi(\hat p,W)$ is a degree-$s$ polynomial in the variable $W$, $\psi^{(\ell)}(\nu,q,Z)$ is degree-$\ell$ polynomial in the variable $Z$. 
	Even though they are not normalisable, a similar decomposition can be written in terms of basis states $|p,W\rangle\equiv |p\rangle_{\mu_1...\,\mu_s}W^{\mu_1}...\,W^{\mu_s}$ and $|\nu,q,Z\rangle\equiv |\nu,\vec x\rangle_{\mu_1...\,\mu_\ell} Z^{\mu_1}...\, Z^{\mu_\ell}$, namely
	\begin{equation}
	\label{spinning state decomposition}
	|p,W\rangle=\sum_{\ell=0}^s \frac{1}{\ell!(h-1)_\ell} \int_0^\infty d\nu\, \mu_{s,\ell}(\nu) \int d^d\vec x\, G^{(s)}_{-\nu,\ell}(\hat p,W;q,D_Z)\, |\nu,q,Z\rangle\,,
	\end{equation}
	with inverse
	\begin{equation}
	\label{nu,x spinning}
	|\nu,q,Z \rangle=\frac{1}{s!\left(\frac{d-1}{2} \right)_s}  \int [d^{d+1}\hat p]\, G^{(s)}_{\nu,\ell}(\hat p,K_W;q,Z)\, |p,W\rangle\,.
	\end{equation}
	Written with tensor indices, these equations amount to
	\begin{equation}
	|p\rangle_{a_1...\,a_s}=\sum_{\ell=0}^s \int_0^\infty d\nu\, \mu_{s,\ell}(\nu) \int d^d\vec x\, [G^{(s)}_{-\nu,\ell}(\hat p;\vec x)]_{a_1...\,a_s}^{i_1...\,i_\ell}\, |\nu,\vec x\rangle_{i_1...\,i_\ell}\,,
	\end{equation}
	and
	\begin{equation}
	|\nu,\vec x \rangle_{i_1...\,i_\ell}= \int [d^{d+1}\hat p]\, [G^{(s)}_{\nu,\ell}(\hat p;\vec x)]_{i_1...\,i_\ell}^{a_1...\,a_s}\, |p\rangle_{a_1...\,a_s}\,.
	\end{equation}
	
	A generic one-particle state \eqref{generic massive state} can now be decomposed, using \eqref{psi(p,W) decomposition}-\eqref{nu,x spinning} together with the completeness relation \eqref{Costa et al},\footnote{Alternatively, one should be able to use the orthogonality relation given in Eq.~(227) of \cite{Costa:2014kfa}.}
	\begin{equation}
	\label{generic psi decomposition tensor}
	\begin{split}
	|\psi\rangle&=\frac{1}{s!\left(\frac{d-1}{2} \right)_s} \int [d^{d+1}\hat p]\, \psi(\hat p,K_W) |p,W\rangle\\
	&=\sum_{\ell=0}^s \frac{1}{\ell!(h-1)_\ell} \int_0^\infty d\nu\, \mu_{s,\ell}(\nu) \int d^d\vec x\,\psi^{(\ell)}(\nu,q,D_Z)|\nu,q,Z\rangle\,.
	\end{split}
	\end{equation}
	As before, one can check that the Lorentz inner product \eqref{Lorentz inner product} is recovered from the particle inner product \eqref{massive inner product} together with the decomposition \eqref{generic psi decomposition tensor}. Using the same set of identities, we can indeed write
	\begin{equation}
	\begin{split}
	\langle \phi|\psi\rangle
	&=\frac{1}{s!\left(\frac{d-1}{2} \right)_s} \int [d^{d+1}\hat p]\, \phi(\hat p,K_W)^* \psi(\hat p,W)\\
	&=\sum_{\ell=0}^s \frac{1}{\ell!(h-1)_\ell} \int_0^\infty d\nu\, \mu_{s,\ell}(\nu) \int d^d\vec x\, \phi^{(\ell)}(\nu,q,D_Z)^*\, \psi^{(\ell)}(\nu,q,Z)\,,
	\end{split}
	\end{equation}
	as expected from \eqref{Lorentz inner product}.
	
	Let us show that the states \eqref{nu,x spinning} indeed transform by the Lorentz representations of the spinning principal series described in section~\ref{section 2}. Acting with a Lorentz transformation $U(\Lambda)$, we have
	\begin{equation}
	U(\Lambda)|\nu,q,Z \rangle=\frac{1}{s!\left(\frac{d-1}{2} \right)_s} \int [d^{d+1}\hat p]\, G^{(s)}_{\nu,\ell}(\hat p,K_W;q,Z)\, U(\Lambda) |p,W\rangle\,.
	\end{equation}
	At this point we need to specify the transformation $U(\Lambda) |p,W\rangle$ corresponding to the standard transformation \eqref{abstract Wigner} of the massive particle states. We claim that it is simply given by
	\begin{equation}
	\label{U |p,W)}
	U(\Lambda)|p,W\rangle=|\Lambda p,\Lambda W\rangle\,,
	\end{equation}
	which in components reads
	\begin{equation}
	\label{transfo |p>mu}
	U(\Lambda)|p\rangle_{\mu_1...\,\mu_s}=\Lambda\indices{^{\nu_1}_{\mu_1}}...\, \Lambda\indices{^{\nu_s}_{\mu_s}} |\Lambda p\rangle_{\nu_1...\,\nu_s}\,.
	\end{equation}
	Let us prove that \eqref{transfo |p>mu} induces the correct transformation of massive vector particles $|p\rangle_a$, which straightforwardly generalises to higher integer spin. Using \eqref{h from H}, we can write
	\begin{equation}
	\begin{split}
	U(\Lambda)|p\rangle_a&=E(\hat p)^\mu_a\, U(\Lambda)|p\rangle_\mu=E(\hat p)^\mu_a \Lambda\indices{^\nu_\mu} |\Lambda p \rangle_\nu\\
	&=E(\hat p)^\mu_a \Lambda\indices{^\nu_\mu} \left(-(\Lambda \hat p)_\nu (\Lambda \hat p)^\sigma +E^{-1}(\Lambda \hat p)^b_\nu E(\Lambda \hat p)_b^\sigma  \right)|\Lambda p\rangle_\sigma\\
	&=E^{-1}(\Lambda \hat p)_\nu^b\, \Lambda\indices{^\nu_\mu} E(\hat p)^\mu_a |\Lambda p\rangle_b-E(\hat p)^\mu_a \Lambda\indices{^\nu_\mu} (\Lambda \hat p)_\nu\, (\Lambda \hat p)^\sigma |\Lambda p\rangle_\sigma\,,
	\end{split}
	\end{equation}
	where we inserted the resolution of the identity \eqref{resolution of identity E} in writing the second line. The last term actually vanishes on account of the transversality $\hat p^\mu |p\rangle_\mu=0$ of the ambient vector $|p\rangle_\mu$. We are therefore left with
	\begin{equation}
	\label{Wigner transfo}
	U(\Lambda)|p\rangle_a=D(\Lambda,\hat p)\indices{^b_a} |\Lambda p\rangle_b\,,
	\end{equation}
	where we defined the matrix 
	\begin{equation}
	D(\Lambda,\hat p)\indices{^b_a}\equiv E^{-1}(\Lambda \hat p)_\nu^b\, \Lambda\indices{^\nu_\mu} E(\hat p)^\mu_a\,. 
	\end{equation}
	Equation \eqref{Wigner transfo} is exactly the transformation \eqref{abstract Wigner} of a massive Wigner UIR, with $D(\Lambda,\hat p)\indices{^b_a}$ the explicit representation of the Wigner rotation \eqref{Wigner rotation abstract} leaving the reference momentum \eqref{kmu} invariant. One can check that $E(\hat p)^\mu_a=L(\hat p)\indices{^\mu_a}$ where $L(\hat p)\indices{^\mu_\nu}$ is the boost matrix \eqref{boost matrix} relating the generic momentum $p^\mu$ to the the reference vector $k^\mu$. This demonstrates the validity of the transformation \eqref{U |p,W)}. In appendix~\ref{app:1}, we show that \eqref{U |p,W)} also induces the correct Wigner rotations in the case of massless particles.
	
	Returning to the Lorentz transformation of the states $|\nu,q,Z\rangle$, we thus have
	\begin{equation}
	\begin{split}
	U(\Lambda)|\nu,q,Z \rangle&=\frac{1}{s!\left(\frac{d-1}{2} \right)_s} \int [d^{d+1}\hat p]\, G^{(s)}_{\nu,\ell}(\hat p,K_W;q,Z)\, |\Lambda p,\Lambda W\rangle\\
	&=\frac{1}{s!\left(\frac{d-1}{2} \right)_s}  \int [d^{d+1}\hat p]\, G^{(s)}_{\nu,\ell}(\Lambda^{-1} \hat p,\Lambda^{-1} K_{W'};q,Z)]\, |p,W'\rangle\,.
	\end{split}
	\end{equation}
	where we used the Lorentz invariance of the integral measure, and we renamed $W'=\Lambda W$ as well as used the explicit expression \eqref{K mu} to derive
	\begin{equation}
	(K_W)_\mu=\Lambda\indices{^\nu_\mu} (K_{W'})_\nu\,, \qquad K_W=\Lambda^{-1} K_{W'}\,.
	\end{equation}
	From \eqref{spinning intertwiner} and the relation \eqref{relation}, we can show that 
	\begin{equation}
	\begin{split}
	G^{(s)}_{\nu,\ell}(\Lambda^{-1} \hat p,\Lambda^{-1} K_W;q(\vec x),Z)&=G^{(s)}_{\nu,\ell}(\hat p, K_W;\Lambda q(\vec x),\Lambda Z)\\
	&=\Omega(x)^{\Delta_\nu}\, G^{(s)}_{\nu,\ell}(\hat p,K_W;q(\vec x\, '),\Lambda Z)\,,
	\end{split}
	\end{equation}
	with $\vec x\,'=\Lambda \vec x$, such that
	\begin{equation}
	\label{U|q,Z)}
	U(\Lambda) |\nu, q(\vec x),Z\rangle
	= \Omega(x)^{\Delta_\nu}\, |\nu,q(\vec x\,') ,\Lambda Z\rangle\,.
	\end{equation}
	In components this equation reads
	\begin{equation}
	U(\Lambda) |\nu, \vec x\rangle_{\mu_1...\,\mu_\ell}
	= \Omega(x)^{\Delta_\nu}\, \Lambda\indices{^{\nu_1}_{\mu_1}}...\, \Lambda\indices{^{\nu_s}_{\mu_s}} |\nu,\vec x\, '\rangle_{\nu_1...\,\nu_\ell}\,.
	\end{equation}
	Now we apply \eqref{pullback} in order to extract the transformation of the physical states. We have
	\begin{equation}
	\label{U|nu,x)}
	\begin{split}
	U(\Lambda) |\nu, \vec x\rangle_{i_1...\,i_\ell}&= \frac{\partial q^{\mu_1}}{\partial x^{i_1}}...\, \frac{\partial q^{\mu_\ell}}{\partial x^{i_\ell}}\, U(\Lambda) |\nu, \vec x\rangle_{\mu_1...\,\mu_\ell}\\
	&= \Omega(x)^{\Delta_\nu}\, \Lambda\indices{^{\nu_1}_{\mu_1}} \frac{\partial q^{\mu_1}}{\partial x^{i_1}}...\, \Lambda\indices{^{\nu_s}_{\mu_s}} \frac{\partial q^{\mu_\ell}}{\partial x^{i_\ell}} |\nu,\vec x\, '\rangle_{\nu_1...\,\nu_\ell}\,.
	\end{split}
	\end{equation}
	Differentiating \eqref{relation}, we find
	\begin{equation}
	\label{partial q'}
	\Lambda\indices{^\mu_\nu} \frac{\partial q^\nu(\vec x)}{\partial x^i}=\Omega(x)^{-1} \frac{\partial x'^j}{\partial x^i} \frac{\partial q^\mu(\vec x\, ')}{\partial x'^j}+ \frac{\partial}{\partial x^i}\left( \Omega(x)^{-1}\right) q^\mu(\vec x\,')\,.
	\end{equation}
	The second term does not contribute to \eqref{U|nu,x)} thanks to the transversality $q^\mu (\vec x\,') |\nu,\vec x\,'\rangle_{\mu...\,\mu_\ell}=0$ of the ambient tensor $|\nu,\vec x\,'\rangle_{\mu_1...\mu_\ell}=|\nu,q(\vec x\,')\rangle_{\mu_1...\mu_\ell}$.
	Hence
	\begin{equation}
	\label{transfo spinning conformal states}
	\begin{split}
	U(\Lambda) |\nu, \vec x\rangle_{i_1...\,i_\ell}&=\Omega(x)^{\Delta_\nu-\ell}\, \frac{\partial x'^{j_1}}{\partial x^{i_1}} \frac{\partial q^{\mu_1}(\vec x\,')}{\partial x'^{j_1}}...\,\frac{\partial x'^{j_\ell}}{\partial x^{i_\ell}} \frac{\partial q^{\mu_\ell}(\vec x\,')}{\partial x'^{j_\ell}} |\nu,\vec x\, '\rangle_{\mu_1...\,\mu_\ell}\\
	&=\Omega(x)^{\Delta_\nu-\ell}\, \frac{\partial x'^{j_1}}{\partial x^{i_1}}...\,\frac{\partial x'^{j_\ell}}{\partial x^{i_\ell}} |\nu,\vec x\, '\rangle_{j_1...\,j_\ell}\\
	&=\Omega(x)^{\Delta_\nu}\, R\indices{^{j_1}_{i_1}}(x)\,...\,R\indices{^{j_\ell}_{i_\ell}}(x) |\nu,\vec x\, '\rangle_{j_1...\,j_\ell}\,,
	\end{split}
	\end{equation}
	using \eqref{finite conformal transfo} in the last equality.
	As expected we recover the transformation \eqref{Lorentz finite transfo} of a Lorentz UIR $[\Delta_\nu,\ell]$ of the spinning principal series.
	
	\subsection{Half spin}
	Finally we turn to the case of massive particles with half-integer spin. These are characterised by a spinor representation of the little group $\operatorname{SO}(d+1)$, i.e., a unitary representation of $\operatorname{Spin}(d+1)$. As for tensors, it is useful to embed these spinors of $\operatorname{SO}(d+1)$ into Lorentz-covariant spinors of $\operatorname{SO}(1,d+1)$. The canonical way of describing Lorentz-covariant spinors is to introduce the Dirac matrices $\Gamma_\mu$ of dimension $2^{[\frac{d}{2}]+1}$ that satisfy the anticommutation relations
	\begin{equation}
	\Gamma_\mu \Gamma_\nu+\Gamma_\nu \Gamma_\mu=2\eta_{\mu\nu} \mathds{1}\,.
	\end{equation}
	The matrices $\Gamma_a$ are hermitian, while $\Gamma_0$ is anti-hermitian. The Lorentz generators in this representation are then defined as
	\begin{equation}
	\label{Dirac representation}
	\Sigma_{\mu\nu}=\frac{i}{4} [\Gamma_\mu,\Gamma_\nu]\,.
	\end{equation}
	In this way we can define spinor wavefunctions $\Psi(\hat p)$ which are Dirac spinors of $\operatorname{SO(1,d+1)}$ of dimension $2^{[\frac{d}{2}]+1}$, i.e.,  under Lorentz transformation they behave like
	\begin{equation}
	\label{Lorentz spinor transfo}
	U(\Lambda)\Psi(\hat p)=S(\Lambda)^{-1}\, \Psi(\Lambda \hat p)\,,
	\end{equation}
	with $S(\Lambda)$ the Dirac spinor representation of $\operatorname{SO}(1,d+1)$.
	However these are not irreducible spinors with respect to the massive little group $\operatorname{SO}(d+1)$ generated by $\langle J_{ab} \rangle$. Since the matrix $\Gamma_0$ commutes with all the rotation matrices $\Sigma_{ab}$ and further satisfies $(\Gamma_0)^2=-\mathds{1}$, it can be used to define the projectors
	\begin{equation}
	\Pi_\pm(\hat k)\equiv \frac{1}{2}(\mathds{1} \pm i\Gamma_0)\,,
	\end{equation}
	which project onto the subspaces corresponding to irreducible spinor representations of $\operatorname{SO}(d+1)$, each one of dimension $2^{[\frac{d}{2}]}$. In particular one can diagonalise $\Gamma_0$ and group its eigenspinors according to their eigenvalue,
	\begin{equation}
	\Gamma_0\, u_\sigma^+(\hat k)=-i u_\sigma^+(\hat k)\,, \qquad \Gamma_0\, u^-_\sigma(\hat k)=i u^-_\sigma(\hat k)\,,
	\end{equation}
	with the index running from $\sigma=1,...\,,\left[\frac{d}{2}\right]$. This obviously implies
	\begin{equation}
	\Pi_\pm(\hat k)\, u^\pm_\sigma(\hat k)=u^\pm_\sigma(\hat k)\,, \qquad \Pi_\pm(\hat k)\, u^\mp_\sigma(\hat k)=0\,. 
	\end{equation}
	The above equations, valid at the reference momentum $\hat k$, can be generalised to an arbitrary momentum $\hat p$ by applying the spinor representation $S(L(\hat p))$ of the boost element $L(\hat p)$, and introducing
	\begin{align}
	u^\pm_\sigma(\hat p)&=S(L(\hat p))\, u^\pm_\sigma(\hat k)\,,\\
	\Pi_\pm(\hat p)&=S(L(\hat p))\, \Pi_\pm(\hat k)\, S(L(\hat p))^{-1}=\frac{1}{2}(\mathds{1}\pm i \hat p^\mu \Gamma_\mu)\,.
	\end{align}
	Introducing the dual spinors $\bar u^\pm_\sigma(\hat k)$ and $\bar u^\pm_\sigma(\hat p)$ satisfying  
	\begin{equation}
	\bar u^\pm_\sigma(\hat p) u^\pm_{\sigma '} (\hat p)=\bar u^\pm_\sigma(\hat k) u^\pm_{\sigma '} (\hat k)=\delta_{\sigma \sigma'}\,, \qquad \bar u^\pm_\sigma(\hat p) u^\mp_{\sigma '} (\hat p)=\bar u^\pm_\sigma(\hat k) u^\mp_{\sigma '} (\hat k)=0\,,
	\end{equation}
	as well as
	\begin{equation}
	\bar u^\pm_\sigma(\hat p)=\bar u^\pm_\sigma(\hat k) S(L(\hat p))^{-1}\,,
	\end{equation}
	we can give a resolution of the projectors $\Pi_\pm(\hat p)$,  
	\begin{equation}
	\Pi_\pm(\hat p)= u^{\pm\sigma}(\hat p)\, \bar u_\sigma^{\pm}(\hat p)\,, 
	\end{equation}
	where $\sigma$ indices are raised and lowered with the Kronecker delta $\delta_{\sigma \sigma'}$.
	The dual spinors $\bar u^\pm_\sigma(\hat p)$ can be used to project onto the irreducible $\operatorname{SO}(d+1)$ spinors transforming according to Wigner's induced representations, defined as
	\begin{equation}
	\label{Wigner spinors}
	\psi^\pm_\sigma(\hat p) \equiv \bar u^\pm_\sigma(\hat p) \Psi(\hat p)\,.
	\end{equation}
	Indeed, given the Lorentz transformation \eqref{Lorentz spinor transfo} of the Dirac spinor $\Psi(\hat p)$, for the wavefunctions \eqref{Wigner spinors} we have
	\begin{equation}
	\begin{split}
	U(\Lambda)\psi^\pm_\sigma(\hat p) &= \bar u^\pm_\sigma(\hat p) S(\Lambda)^{-1} \Psi(\Lambda \hat p)\\
	&=\bar u^\pm_\sigma(\hat p) S(\Lambda)^{-1} \left(u^{+\rho}(\Lambda \hat p)\, \bar u_\rho^{+}(\Lambda \hat p)+u^{-\rho}(\Lambda \hat p)\, \bar u_\rho^{-}(\Lambda \hat p)\right)\Psi(\Lambda \hat p)\\
	&=\bar u^\pm_\sigma(\hat p) S(\Lambda)^{-1}\, u^{\pm\rho}(\Lambda \hat p)\, \bar u_\rho^{\pm}(\Lambda \hat p) \Psi(\Lambda \hat p)\\
	&=\bar u^\pm_\sigma(\hat k) S(L(\hat p))^{-1} S(\Lambda)^{-1} S(L(\Lambda \hat p))\, u^{\pm \rho}(\hat k)\, \psi^\pm_\rho(\Lambda \hat p)\\
	&=\bar u^\pm_\sigma(\hat k) S(L^{-1}(\Lambda \hat p) \Lambda L(\hat p))^{-1}\, u^{\pm \rho}(\hat k)\, \psi^\pm_\rho(\Lambda \hat p)\\
	&=[D^\pm(W(\Lambda,\hat p))^{-1}]\indices{_\sigma^{\sigma'}} \psi^\pm_{\sigma'}(\Lambda \hat p)\,,
	\end{split}
	\end{equation}
	where $W(\Lambda,\hat p) \in \operatorname{SO}(d+1)$ is the little group element \eqref{Wigner rotation abstract}, and 
	\begin{equation}
	[D^\pm(W(\Lambda,\hat p))^{-1}]\indices{_\sigma^{\sigma'}}\equiv \bar u^\pm_\sigma(\hat k) S(W(\Lambda,\hat p))^{-1}\, u^{\pm \sigma'}(\hat k)\,,
	\end{equation}
	are the components of one of its unitary irreducible spinor representation. To summarise, a Lorentz spinor $\Psi(\hat p)$ behaving like \eqref{Lorentz spinor transfo} breaks into two irreducible spinor particle representations $\psi^\pm_\sigma(\hat p)$. Of course up to this point this is just a recap of the well-known theory of Dirac. 
	
	As in the case of massive particles with integer spin, the use of the Lorentz-covariant spinors $\Psi(\hat p)$ will facilitate the decomposition into elementary representations of Lorentz. The resolution of the identity we need has been given by one of us in \cite{Iacobacci:2020por},
	\begin{equation}
	\int_0^\infty d\nu\, \mu_{\frac{1}{2}}(\nu) \int d^d\vec x\, \frac{\Pi_\pm(\hat p_1) q^\mu(\vec x) \Gamma_\mu\, \Pi_\pm(\hat p_2)}{(-\hat p_1 \cdot q(\vec x))^{\Delta_\nu+\frac{1}{2}} (-\hat p_2 \cdot q(\vec x))^{\Delta^*_\nu+\frac{1}{2}}}=\mp i\, \delta(\hat p_1-\hat p_2) \Pi_\pm(\hat p_2)\,,
	\label{complSpinor}
	\end{equation}
	with
	\begin{equation}
	\mu_{\frac{1}{2}}(\nu)=\frac{1}{2\pi^{d+1}}\frac{\Gamma(\frac{d+1}{2}+i\nu)\Gamma(\frac{d+1}{2}-i\nu)}{\Gamma(\frac{1}{2}+i\nu)\Gamma(\frac{1}{2}-i\nu)}\,.
	\end{equation}
	The corresponding orthogonality relation in this case is obtained from appendix C of \cite{Iacobacci:2020por},
	\begin{equation}
	\begin{split}
	&\pm 2i \int [d^{d+1}\hat p]\,
	\frac{\begin{pmatrix}
		-1 & \gamma^i x_i  
		\end{pmatrix} \Pi_\pm(\hat p) \begin{pmatrix}
		\gamma^i x'_i \\
		1
		\end{pmatrix}}
	{(-\hat p \cdot q(\vec x))^{\Delta^*_\nu+\frac{1}{2}}(-\hat p \cdot q(\vec x\,'))^{\Delta_{\nu'}+\frac{1}{2}}}=\frac{\delta(\nu -\nu')\delta^{(d)}(\vec{x}-\vec{x}\,')}{\mu_{\frac{1}{2}}(\nu)}\,,
	\end{split}
	\end{equation}
	for $\nu,\nu' \in \mathbb{R}^+$.
	The presence of the projector $\Pi_\pm(\hat p)$ indicates that the resolution of the identity \eqref{complSpinor} is within the subspace of Lorentz-covariant Dirac spinors that correspond to a single set of Wigner particles. Acting with \eqref{complSpinor} on a generic spinor wavefunction $\Psi(\hat p_2)$ and integrating over $\hat p_2$ thus yields
	\begin{equation}
	\label{Spinor decomposition}
	\Pi_\pm(\hat p) \Psi(\hat p)=\pm i\, \Pi_\pm(\hat p) \int_0^\infty d\nu\, \mu_{\frac{1}{2}}(\nu) \int d^d\vec x\, \frac{q^\mu(\vec x) \Gamma_\mu \Psi_\pm (\nu,\vec x)}{(-\hat p \cdot q(\vec x))^{\Delta_\nu+\frac{1}{2}}} \,,
	\end{equation}
	with
	\begin{equation}
	\label{Spinor decomposition bis}
	\Psi_\pm(\nu,\vec x)=\int [d^{d+1}\hat p]\,\frac{\Pi_\pm(\hat p) \Psi(\hat p)}{(-\hat p \cdot q(\vec x))^{\Delta^*_\nu+\frac{1}{2}}} \,.
	\end{equation}
	This decomposition of the particle wavefunction $\Pi_\pm(\hat p) \Psi(\hat p)$ is completely analogous to the spinless case described in \eqref{wavefunction decomposition}-\eqref{wavefunction decomposition bis} up to the shift $\Delta_\nu \mapsto \Delta_\nu+\frac{1}{2}$. The qualitatively new ingredient is the presence of the operator $q^\mu(\vec x) \Gamma_\mu$ in \eqref{Spinor decomposition}, whose role is to ensure that $q^\mu(\vec x) \Gamma_\mu \Psi_\pm(\nu,\vec x)$ carries an elementary representation of the Lorentz group. To understand this point better, let us focus on the locus $\vec x=0$ where the inducing representation \eqref{inducing representation} in the construction of elementary representation of Lorentz is defined. In particular within an elementary representation, the generator $D=-J_{0(d+1)}$ is proportional to the identity. This is not the case of the Dirac representation $\Sigma_{0(d+1)}=\frac{i}{4}[\Gamma_0,\Gamma_{d+1}]$, and we should therefore project onto the subspaces of spinors that have common eigenvalue with respect to $\Sigma_{0(d+1)}$. This is achieved with the projectors 
	\begin{equation}
	\label{P(0)}
	P_\pm(0)=\frac{1}{2}\left(\mathds{1}\pm 2i\, \Sigma_{0(d+1)} \right)\,,
	\end{equation}
	which project onto subspaces that carry spinor representations of $\operatorname{SO}(d)$ of dimension $2^{[\frac{d}{2}]}$, the correct dimension for Dirac spinors of $\operatorname{SO}(d)$. To be very explicit, let us adopt the following representation of the gamma matrices $\Gamma_\mu$,
	\begin{equation}
	\Gamma_i=\begin{pmatrix}
	\gamma_i & 0\\
	0 & -\gamma_i
	\end{pmatrix}\,, \qquad \Gamma_0=\begin{pmatrix}
	0 & -1\\
	1 & 0
	\end{pmatrix}\,, \qquad \Gamma_{d+1}=\begin{pmatrix}
	0 & 1\\
	1 & 0
	\end{pmatrix}\,,
	\end{equation}
	in terms of the lower-dimensional gamma matrices $\gamma_i$ (and we denote the $2^{[\frac{d}{2}]} \times 2^{[\frac{d}{2}]}$ unit matrix by 1). Within this representation, the rotations $\Sigma_{ij}$ and boost matrix $\Sigma_{0(d+1)}$ take the explicit form
	\begin{equation}
	\Sigma_{ij}=\frac{i}{4}\begin{pmatrix}
	[\gamma_i,\gamma_j] & 0\\
	0 & [\gamma_i,\gamma_j]
	\end{pmatrix}\,, \qquad \Sigma_{0(d+1)}=-\frac{i}{2} \begin{pmatrix}
	1 & 0\\
	0 & -1
	\end{pmatrix}\,.
	\end{equation}
	Thus, if we write a generic Dirac spinor of $\operatorname{SO}(1,d+1)$ as
	\begin{equation}
	\Psi(0)=\begin{pmatrix}
	\psi_+(0)\\ \psi_-( 0)    
	\end{pmatrix}\,,
	\end{equation}
	we see that $\psi_\pm(0)$ are Dirac spinors of $\operatorname{SO}(d)$, and that the role of \eqref{P(0)} is precisely to project onto the subspaces they span,
	\begin{equation}
	P_+(0)\, \Psi(0)=\begin{pmatrix}
	\psi_+(0)\\ 0    
	\end{pmatrix}\,, \qquad 
	P_-(0)\, \Psi(0)=\begin{pmatrix}
	0\\ \psi_-(0)
	\end{pmatrix}\,.
	\end{equation}
	One can exctract the Dirac spinors of $\operatorname{SO}(d)$ by using the basis $v_\pm^\sigma(0)$ of Lorentz-covariant spinors, grouped into the following matrices
	\begin{equation}
	v_+(0)=\begin{pmatrix}
	1 \\ 0  
	\end{pmatrix}\,, \qquad v_-(0)=\begin{pmatrix}
	0 \\ 1 
	\end{pmatrix}\,,
	\end{equation}
	with duals
	\begin{equation}
	\bar v_+(0)=\begin{pmatrix}
	1 & 0    
	\end{pmatrix}\,, \qquad \bar v_-(0)=\begin{pmatrix}
	0 & 1    
	\end{pmatrix}\,,
	\end{equation}
	such that
	\begin{equation}
	\bar v_\pm (0) \Psi(0)=\psi_\pm(0)\,.
	\end{equation}
	To get the full structure \eqref{inducing representation} of the elementary representation at $\vec x=0$, we should also check that the generators $K_i=J_{0i}+J_{(d+1)i}$ act trivially. Their representation on Lorentz-covariant spinors is given by
	\begin{equation}
	K^{(\Gamma)}_i\equiv \Sigma_{0i}+\Sigma_{(d+1)i}=i\begin{pmatrix}
	0 & 0\\
	\gamma_i & 0
	\end{pmatrix}\,,
	\end{equation}
	such that its action on a generic spinor $\Psi(0)$ yields
	\begin{equation}
	K^{(\Gamma)}_i\, \Psi(0)=\begin{pmatrix}
	0 \\ i\gamma_i\, \psi_+(0)    
	\end{pmatrix}\,.
	\end{equation}
	Since only $P_-(0) \Psi(0)$ is annihilated by $K_i^{(\Gamma)}$, it is the only component of $\Psi(0)$ which can be associated with an elementary representation of Lorentz. We can now `translate' the above spinors to a generic point $\vec x$ by application of the operators $e^{-i x^i P_i}$ in the Lorentz-covariant spinor representation. We have
	\begin{equation}
	P_i^{(\Gamma)}\equiv \Sigma_{0i}-\Sigma_{(d+1)i}=i\begin{pmatrix}
	0 & \gamma_i\\
	0 & 0
	\end{pmatrix}\,,
	\end{equation}
	such that the latter is given by
	\begin{equation}
	e^{-i x^i P^{(\Gamma)}_i}=\mathds{1}-i x^i P^{(\Gamma)}_i=\begin{pmatrix}
	1 &  x^i\gamma_i\\
	0 & 1
	\end{pmatrix}\,.
	\end{equation}
	We therefore have the translated basis spinors
	\begin{equation}
	v_+(\vec x)\equiv e^{-i x^i P_i^{(\Gamma)}} v_+(0)=\begin{pmatrix}
	1 \\ 0  
	\end{pmatrix}\,, \qquad v_-(\vec x)\equiv e^{-i x^i P_i^{(\Gamma)}}v_-(0)=\begin{pmatrix}
	x^i \gamma_i \\ 1 
	\end{pmatrix}\,, 
	\end{equation}
	with dual spinors
	\begin{equation}
	\bar v_+(\vec x)=\begin{pmatrix}
	1 & -x^i \gamma_i    
	\end{pmatrix}\,, \qquad \bar v_-(\vec x)=\begin{pmatrix}
	0 & 1    
	\end{pmatrix}\,.
	\end{equation}
	An explicit computation then shows that the quantity $q^\mu(\vec x) \Gamma_\mu$ of interest can be simply written as 
	\begin{equation}
	q^\mu(\vec x) \Gamma_\mu=2 \begin{pmatrix}
	x^i \gamma_i & -x^2\\
	1 & -x^i\gamma_i
	\end{pmatrix}=2 v_-(\vec x) \bar v_+(\vec x)\,.
	\end{equation}
	Applied on a generic Dirac spinor 
	\begin{equation}
	\Psi(\vec x)=\begin{pmatrix}
	\psi_1(\vec x)\\ \psi_2(\vec x)    
	\end{pmatrix}
	\end{equation}
	it thus yields
	\begin{equation}
	q^\mu(\vec x) \Gamma_\mu \Psi(\vec x)=2 v^\sigma_-(\vec x) \left[\bar v_+(\vec x) \Psi(\vec x)\right]_\sigma=2 v^\sigma_-(\vec x) \left[\psi_1(\vec x)-x^i \gamma_i\, \psi_2(\vec x)\right]_\sigma\,,
	\end{equation}
	where we reinstated $\sigma$ indices for clarity. 
	On the one hand, the above quantity is a linear combination of the Dirac spinors $v_-^\sigma(\vec x)$ that are annihilated by $K_i^{(\Gamma)}$ at $\vec x=0$. On the other hand, it is proportional to the $\operatorname{SO}(d)$ Dirac spinor wavefunction $\psi_1(\vec x)-x^i \gamma_i\, \psi_2(\vec x)$ which, provided $\Psi(\vec x)$ transforms like a Lorentz-covariant conformal spinor (see \eqref{conformal Lorentz spinor} below), is known to be the only combination of $\psi_1(\vec x)$ and $\psi_2(\vec x)$ which carries an elementary representation of the Lorentz group \cite{Isono:2017grm}. With this understanding at hand, we rewrite the decomposition \eqref{Spinor decomposition} in the slightly more transparant way, 
	\begin{equation}
	\label{Spinor decomposition v2}
	\Pi_\pm(\hat p) \Psi(\hat p)=\pm 2i\, \Pi_\pm(\hat p) \int_0^\infty d\nu\, \mu_{\frac{1}{2}}(\nu) \int d^d\vec x\, \frac{v^\sigma_-(\vec x) [\bar v_+(\vec x) \Psi_{\pm} (\nu,\vec x)]_\sigma}{(-\hat p \cdot q(\vec x))^{\Delta_\nu+\frac{1}{2}}} \,.
	\end{equation}
	These equations express the decomposition of the Lorentz-covariant Dirac spinors $\Pi_\pm(\hat p) \Psi(\hat p)$ in terms of $\bar v_+(\vec x) \Psi_{\pm} (\nu,\vec x)$ which are Dirac spinors of $\operatorname{SO}(d)$. Let us now complete the proof that these transform according to elementary representations of the principal continuous series. Starting from 
	\begin{equation}
	U(\Lambda)\Pi_\pm(\hat p) \Psi(\hat p)=S(\Lambda)^{-1} \Pi_\pm(\Lambda \hat p) \Psi(\Lambda \hat p)\,,
	\end{equation}
	we have
	\begin{equation}
	\label{conformal Lorentz spinor}
	\begin{split}
	U(\Lambda) \Psi_\pm(\nu,\vec x)&=S(\Lambda)^{-1} \int [d^{d+1}\hat p]\,\frac{\Pi_\pm(\Lambda \hat p) \Psi(\Lambda \hat p)}{(-\hat p \cdot q(\vec x))^{\Delta^*_\nu+\frac{1}{2}}}\\
	&=S(\Lambda)^{-1} \int [d^{d+1}\hat p]\,\frac{\Pi_\pm(\hat p) \Psi(\hat p)}{(-\hat p \cdot \Lambda q(\vec x))^{\Delta^*_\nu+\frac{1}{2}}}\\
	&=\Omega(x)^{\Delta_\nu^*+\frac{1}{2}}\, S(\Lambda)^{-1} \Psi_\pm(\nu,\vec x)\,.\\
	\end{split}
	\end{equation}
	This is the `conformal Lorentz-covariant spinor transformation' mentioned above. Calling the result obtained by Isono \cite{Isono:2017grm}, we thus have 
	\begin{equation}
	U(\Lambda) [\bar v_+(\vec x) \Psi_\pm(\nu,\vec x)]=\Omega(x)^{\Delta_\nu^*}\, \tilde S(R(\vec x))^{-1} [\bar v_+(\vec x\, ') \Psi_\pm(\nu,\vec x\, ')]\,,
	\end{equation}
	with $\tilde S(R(x))$ the Dirac spinor representation of $R(x) \in \operatorname{SO}(d)$ given in \eqref{finite conformal transfo}, which is to say that $\bar v_+(\vec x) \Psi_\pm(\nu,\vec x)$ transforms according to the unitary representation $[\Delta_\nu^*,\tilde S]$ of the principal continuous series.\footnote{Note that the Dirac spinor representation of $\operatorname{SO}(d)$ can be further reduced to two Weyl spinors when $d$ is odd.} As in the previous sections, the decomposition formula \eqref{Spinor decomposition v2} is also easily adapted to the basis states instead of the wavefunctions.
	
	\subsection{Half-integer spin}
	While we treated massive spin-$\frac{1}{2}$ particles in the previous subsection, one can in principle generalise to arbitrary half-integer spin $s+\frac{1}{2}$. All one needs to do is to consider the Lorentz-covariant Dirac spinor $\Pi_\pm(\hat p) \Psi(\hat p,W)$ whose components are homogeneous polynomials of degree $s$ in $W^\mu$. One obtains
	\begin{equation}
	\label{tensor spin decomposition}
	\begin{split}
	\Pi_\pm(\hat p)\Psi(\hat p,W)
	&=\pm 2i \Pi_\pm(\hat p)\sum_{\ell=0}^s \frac{1}{\ell!(h-1)_\ell} \int_{0}^{\infty}d\nu\, \mu_{s+\frac{1}{2},\ell}(\nu)\\
	&\times \int d^d \vec x\, G^{(s)}_{\nu,\ell+\frac{1}{2}}(\p,W; q,D_Z)\,  v_-^\sigma(\vec x)  [\bar v_+(\vec x) \Psi^{(\ell)}_\pm (\nu,\vec x,Z)]_\sigma\,,
	\end{split}
	\end{equation}
	with
	\begin{equation}
	\label{tensor spin decomposition bis}
	\Psi^{(\ell)}_\pm(\nu,q,Z)=\frac{1}{s!\left(\frac{d-1}{2} \right)_s} \int [d^{d+1}\hat p]\, G_{-\nu,\ell+\frac{1}{2}}^{(s)}(\p,K_W; q,Z)\, \Pi_\pm(\hat p)\Psi(\hat p,W)\,,
	\end{equation}
	with $\mu_{s+\frac{1}{2},\ell}(\nu)$ a measure left to determine.
	This provides the decomposition of a Lorentz-covariant tensor-spinor $\Pi_\pm(\hat p) \Psi(\hat p,W)$ describing spin-$(s+\frac{1}{2})$ particles, into elementary representations $\Phi_\pm^{(\ell)}(\nu,\vec x,Z)\equiv\bar v_+(\vec x) \Psi^{(\ell)}_\pm (\nu,\vec x,Z)$ carrying tensor-spinor representations of $\operatorname{SO}(d)$. While we are confident that \eqref{tensor spin decomposition}-\eqref{tensor spin decomposition bis} provides the correct decomposition, at this point we have not been able to determine the measure $\mu_{s+\frac{1}{2},\ell}(\nu)$. We hope to resolve this point in the near future, perhaps adapting the determination of $\mu_{s,\ell}(\nu)$ given in appendix D of \cite{Costa:2014kfa}. Note that without further conditions, the tensor-spinors appearing here are reducible. To start with, working with an irreducible set of spin-$(s+\frac{1}{2})$ wavefunctions $\Pi_\pm(\hat p) \Psi(\hat p,W)$ instructs us to supplement the condition $\Gamma^\mu \Psi(\hat p)_{\mu\, ...\, \mu_s}=0$. Second, we need to decompose the conformal tensor-spinor $ \Phi_\pm^{(\ell)}(\nu,\vec x,Z)$ over its $\gamma$-trace and $\gamma$-traceless components \cite{Bure2002SymmetricAO}.
	Defining
	\begin{equation}
	\mathbf{P}\indices{_i^j}=\alpha_\ell\, \gamma_i \gamma^j\,, \qquad    \bar{\mathbf{P}}_i^j=\delta_i^j-\alpha_\ell\, \gamma_i\gamma^j\,, \qquad \alpha_\ell=\frac{\ell}{2(h-1+\ell)}\,,
	\end{equation}
	the $\gamma$-traceless component is explicitly given by
	\begin{equation}
	(\Pi^{(\ell+\frac{1}{2})} \Phi^{(\ell)}_\pm (\nu,\vec x))_{i_1...\,i_\ell}\equiv \frac{1}{\ell!}\, \bar{\mathbf{P}}_{\{i_1}{}^j\Phi_{\,i_2...\, i_\ell\}j}\,,
	\end{equation}
	since it can be shown to satisfy
	\begin{equation}
	\gamma^{i_1}(\Pi^{(\ell+\frac{1}{2})} \Phi^{(\ell)}_\pm (\nu,\vec x))_{i_1...\,i_\ell}=0\,.
	\end{equation}
	Similarly we define the $\gamma$-trace component
	\begin{equation}
	(\Pi^{(\ell-\frac{1}{2})} \Phi^{(\ell)}_\pm (\nu,\vec x))_{i_1...\,i_\ell}\equiv \frac{1}{\ell!}\, \mathbf{P}_{\{i_1}{}^j\Phi_{\,i_2...\, i_\ell\}j}\,,
	\end{equation}
	and we can write the decomposition 
	\begin{equation}
	\Phi^{(\ell)}_\pm (\nu,\vec x)_{i_1...\,i_\ell}=(\Pi^{(\ell+\frac{1}{2})} \Phi^{(\ell)}_\pm (\nu,\vec x))_{i_1...\,i_\ell}+  (\Pi^{(\ell-\frac{1}{2})} \Phi^{(\ell)}_\pm (\nu,\vec x))_{i_1...\,i_\ell} \,.
	\end{equation}
	The wavefunctions $\Pi^{(\ell\pm \frac{1}{2})} \Phi^{(\ell)}_\pm (\nu,\vec x)$ carry the two irreducible spin-$(\ell\pm \frac{1}{2})$ representations of $\operatorname{SO}(d)$ \cite{Bure2002SymmetricAO}. 
	
	\section*{Acknowledgments}
	
	We thank Nicolas Boulanger and Wolfgang Mück for useful discussions. The work of LI is supported by Fondazione Angelo Della Riccia (Florence, Italy). The work of KN is supported by a postdoctoral research fellowship of the
	F.R.S.-FNRS (Belgium).
	
	\appendix
	\section{Massless Wigner rotations}
	\label{app:1}
	In this appendix we use the ambient space formalism to derive the finite form of the transformation of the spin-$\ell$ states $|\nu,\vec x\rangle_\sigma=|\nu,\vec x\rangle_{i_1...\,i_\ell}$ given in \eqref{inverse p decomposition} that appear in the decomposition of massless particle states. Given the particle states $|p\rangle_{i_1...\,i_\ell}$ with $p=\omega q$ as given in \eqref{massless momentum parametrisation}, we introduce the ambient tensor
	\begin{equation}
	|p,Z\rangle=|p\rangle_{\mu_1...\,\mu_\ell}\, Z^{\mu_1}...\,Z^{\mu_\ell}\,,
	\end{equation}
	following the general discussion around \eqref{f(q,Z) definition}. 
	
	We first give the explicit expression of the boost matrix $L(q)\indices{^\mu_\nu}$ relating a generic null momentum $q^\mu(\vec x)$ to the reference momentum
	\begin{equation}
	k^\mu=(1,\vec 0,1)\,.
	\end{equation}
	Introducing a second null vector $n^\mu=(1,\vec 0,-1)$ satisfying $n \cdot q(\vec x)=-2$, we can write its components explicitly as
	\begin{equation}
	\begin{split}
	L(q)\indices{^\mu_0}&=\frac{1}{2}(q^\mu+n^\mu)\,,\\
	L(q)\indices{^\mu_i}&=\frac{1}{2}\frac{\partial q^\mu(\vec x)}{\partial x^i}\,,\\
	L(q)\indices{^\mu_{d+1}}&=\frac{1}{2}(q^\mu-n^\mu)\,,
	\end{split}
	\end{equation}
	and for its inverse we have
	\begin{equation}
	\begin{split}
	L^{-1}(q)\indices{^0_\mu}&=-\frac{1}{2}(q_\mu+n_\mu)\,,\\
	L^{-1}(q)\indices{^i_\mu}&=\frac{1}{2} \delta^{ij} \frac{\partial q_\mu}{\partial x^j}\,,\\
	L^{-1}(q)\indices{^{d+1}_\mu}&=\frac{1}{2}(q_\mu-n_\mu)\,.
	\end{split}
	\end{equation}
	This immediately yields a resolution of the identity in the form 
	\begin{equation}
	\label{massless resolution of 1}
	\delta^\mu_\nu=L(q)\indices{^\mu_\alpha}\, L^{-1}(q)\indices{^\alpha_\nu}=-\frac{1}{2} \left(q^\mu n_\nu+n^\mu q_\nu \right)+\frac{1}{4} \frac{\partial q^\mu}{\partial x^i}  \delta^{ij} \frac{\partial q_\nu}{\partial x^j}\,.
	\end{equation}
	
	With this boost matrix at hand, we can show that the transformation 
	\begin{equation}
	\label{U|p,Z)}
	U(\Lambda)|p,Z\rangle=|\Lambda p,\Lambda Z\rangle\,, \qquad U(\Lambda)|p\rangle_{\mu_1...\,\mu_\ell}=\Lambda\indices{^{\nu_1}_{\mu_1}}...\, \Lambda\indices{^{\nu_\ell}_{\mu_\ell}}|\Lambda p\rangle_{\nu_1...\,\nu_\ell}\,,
	\end{equation}
	is the ambient version of Wigner's transformation \eqref{abstract Wigner}. Let us prove it for a vector representation, also known as photon, for simplicity of notations but without loss of generality. Thus staring from
	\begin{equation}
	U(\Lambda)|p\rangle_\mu=\Lambda\indices{^\nu_\mu}| \Lambda p\rangle_\nu\,,
	\end{equation}
	for the physical states we get
	\begin{equation}
	\begin{split}
	U(\Lambda) |p\rangle_i&=\frac{\partial q^\mu}{\partial x^i} U(\Lambda)|p\rangle_\mu=\frac{\partial q^\mu}{\partial x^i} \Lambda\indices{^\nu_\mu}|\Lambda p \rangle_\nu\\
	&= \frac{\partial q^\mu}{\partial x^i} \Lambda\indices{^\nu_\mu}\left( -\frac{1}{2} \left(\Lambda q^\sigma n_\nu+n^\sigma \Lambda q_\nu \right)+\frac{1}{4} \frac{\partial \Lambda q^\sigma}{\partial x^j}  \delta^{jk} \frac{\partial \Lambda q_\nu}{\partial x^k}\right) |\Lambda p \rangle_\sigma \\
	&=\frac{1}{4} \frac{\partial q^\mu}{\partial x^i} \Lambda\indices{^\nu_\mu}   \delta^{jk} \frac{\partial \Lambda q_\nu}{\partial x^k}|\Lambda p \rangle_j= L^{-1}(\Lambda q)\indices{^j_\nu} \Lambda\indices{^\nu_\mu} L(q)\indices{^\mu_i} |\Lambda p\rangle_j\,,
	\end{split}
	\end{equation}
	where we inserted the resolution of the identity \eqref{massless resolution of 1}, and where we used that $\Lambda q^\sigma |\Lambda p\rangle_\sigma=0$ and $q_\mu \partial_i  q^\mu=0$.
	This is indeed of the form 
	\begin{equation}
	U(\Lambda) |p\rangle_i =D(W(q,\Lambda))\indices{^j_i}|\Lambda p\rangle_j\,,
	\end{equation}
	with 
	\begin{equation}
	D(W(q,\Lambda))\indices{^j_i}=L^{-1}(\Lambda q)\indices{^j_\nu} \Lambda\indices{^\nu_\mu} L(q)\indices{^\mu_i}\,,
	\end{equation}
	the explicit representation of the Wigner rotation \eqref{Wigner rotation abstract}. This demonstrates the validity of the transformation \eqref{U|p,Z)}.
	
	We can then use this to derive the transformation of the states \eqref{inverse p decomposition} under finite Lorentz symmetries. Using again \eqref{relation}, we have
	\begin{equation}
	\begin{split}
	U(\Lambda) |\nu,q(\vec x),Z\rangle&=\frac{1}{\sqrt{2\pi}} \int_0^\infty d\omega\, \omega^{\Delta_\nu-1} U(\Lambda)|\omega q(\vec x),Z\rangle\\
	&=\frac{1}{\sqrt{2\pi}} \int_0^\infty d\omega\, \omega^{\Delta_\nu-1} |\omega \Lambda q(\vec x),\Lambda Z\rangle\\
	&=\frac{1}{\sqrt{2\pi}} \int_0^\infty d\omega\, \omega^{\Delta_\nu-1} |\omega' q(\vec x\,'),\Lambda Z\rangle\\
	&=\frac{\Omega(x)^{\Delta_\nu}}{\sqrt{2\pi}}  \int_0^\infty d\omega'\, \omega'^{\Delta_\nu-1} |\omega' q(\vec x\,'),\Lambda Z\rangle\\
	&=\Omega(x)^{\Delta_\nu}\, |\nu,q(\vec x\,'),\Lambda Z\rangle\,,
	\end{split}     
	\end{equation}
	where we performed the variable change
	\begin{equation}
	\omega'=\Omega(x)^{-1}\, \omega\,.
	\end{equation}
	As shown after \eqref{U|q,Z)}, for the physical components this implies
	\begin{equation}
	\begin{split}
	U(\Lambda) |\nu, \vec x\rangle_{i_1...\,i_\ell}=\Omega(x)^{\Delta_\nu}\, R\indices{^{j_1}_{i_1}}(x)\,...\,R\indices{^{j_\ell}_{i_\ell}}(x) |\nu,\vec x\, '\rangle_{j_1...\,j_\ell}\,,
	\end{split}
	\end{equation}
	as appropriate to a spinning representation of the principal series.
	
	\section{The mixed 3-point amplitude}
	\label{appendix B}
	Here we study the mixed 3-point amplitude \eqref{A3} appearing in the decomposition of the translated massive states. We start by computing the integral 
	\begin{equation}
	\label{3Mixed}
	\mathcal{A}_{\Delta_1, \Delta_2, m}(q_1, q_2; a)= \int [d^{d+1} \hat p]\, \left(-\p\cdot q_1\right)^{\Delta_1}\left(-\p\cdot q_2\right)^{\Delta_2} e^{im \,a^\mu \p_\mu}, \quad (a^\mu \in\mathbb{R}^{1, d+1})\,,
	\end{equation}
	which coincides with \eqref{A3} upon setting $\Delta_1=d/2+i\nu$ and $\Delta_2=d/2-i\nu'$\,.
	To achieve this goal, let us start with the convergent integral
	\begin{equation}
	\label{3MixedConv}
	\mathcal{A}_{\Delta_1, \Delta_2, m}^{\text{conv}}(q_1, q_2; a)=\int [d^{d+1} \hat p]\, \left(-\p\cdot q_1\right)^{\Delta_1}\left(-\p\cdot q_2\right)^{\Delta_2} e^{m\,a\cdot \p}, \quad (a^2<0\,, a^0>0)\,, 
	\end{equation}
	where we assumed $a$ to be timelike and future-directed to ensure  $a\cdot \p<0$, which guarantees the convergence of the integral. We will eventually send $m \to im $ to retrieve the solution of \eqref{3Mixed} inside the lightcone. The solution will then be analytically extended to the other region of $a^\mu \in \mathbb{R}^{1,d+1}$ by inserting an appropriate $i\epsilon$-prescription.
	
	Let us begin our computation by using the Mellin-Barnes representation of the exponential function,
	\begin{equation}
	e^{-x}=\int_{c-i\infty}^{c+i\infty} \frac{d u}{2\pi i}  \,\Gamma(u)\,x^{-u}\,, \qquad c>0,\quad\Re\E(x)>0\,,
	\end{equation}
	to rewrite
	\begin{align}
	\mathcal{A}_{\Delta_1, \Delta_2, m}^{\text{conv}}(q_1, q_2; a)=\int_{-i\infty}^{+i\infty} \frac{d u}{2\pi i}  \Gamma(u)\int [d^{d+1} \hat p] (-\p\cdot q_1)^{\Delta_1}(-\p\cdot q_2)^{\Delta_2}(-m\p\cdot a)^{-u}\,,
	\end{align}
	where the integration contour is chosen to lie to the right of the imaginary axis ($c\to 0^+$). Let us focus on the resolution of
	\begin{equation}
	\label{I integral}
	\mathcal{I}_{\Delta_0,\Delta_1, \Delta_2}(a, q_1,q_2)=\int [d^{d+1} \hat p] (-m\p\cdot a)^{-\Delta_0}(-\p\cdot q_1)^{\Delta_1}(-\p\cdot q_2)^{\Delta_2}\,,
	\end{equation}
	where we set $u\equiv\Delta_0$. Using the Feynman-Schwinger parametrisation, we get
	\begin{equation}
	\mathcal{I}_{\Delta_0,\Delta_1, \Delta_2}(a, q_1,q_2)=\prod_{k=0}^2\int_{0}^{\infty} \frac{d t_k}{t_k}\,\frac{t_k^{\Delta_k}}{\Gamma(\Delta_k)}\int [d^{d+1} \hat p]\,e^{\,\p\cdot T}\,,
	\end{equation}
	with $T=t_0 X+t_1q_1+t_2q_2$ and $X=ma$. We can solve the integral in $\p$ in the rest frame of $T$. Since $T$ is timelike and future-directed, in its rest frame $T^\mu=|T|(1,\vec{0},0)^\mu$, with $|T|=\sqrt{-T^2}$. Then, parametrising $\p^\mu$ with 
	\begin{equation}
	\p^\mu=\left(\frac{1+y^2+|\vec{z}|^2}{2y},  \frac{\vec{z}}{y}, \frac{1-y^2+|\vec{z}|^2}{2y}\right)^\mu,\quad y>0,\quad\vec{z}\in\mathbb{R}^d,
	\end{equation}
	yields
	\begin{equation}
	\begin{split}
	\int [d^{d+1} \hat p]\,e^{\,\p\cdot T}&=\int_{0}^{\infty}\frac{d y}{y^{d+1}}\int d^d z\,e^{-\frac{|T|}{2y}\left(1+y^2+|\vec{z}\,|^2\right)} =\left(\frac{2\pi}{|T|}\right)^{d/2}\int_{0}^{\infty}\frac{d y}{y^{\frac{d+2}{2}}}\,e^{-\frac{|T|}{2y}(1+y^2)}\\
	&\overset{y\to y/2|T|}{=}(4\pi)^{d/2}\int_{0}^{\infty}\frac{d y}{y^{\frac{d+2}{2}}}\,e^{\frac{T^2}{y}-\frac{y}{4}}\,.
	\end{split}
	\end{equation}
	After sending $t_k\to \sqrt{y}\,t_k$, we can compute the integral in $y$,
	\begin{equation}
	\int_{0}^{\infty}\frac{d y}{y^{\frac{d+2}{2}}}\,y^{\frac{\Delta_0+\Delta_1+\Delta_2}{2}}e^{-y/4}=2^{\Sigma\Delta-d}\,\Gamma\left(\frac{\Sigma \Delta-d}{2}\right)\,, \qquad \Sigma \Delta \equiv \Delta_0+\Delta_1+\Delta_2\,,
	\end{equation}
	so that \eqref{I integral} reduces to
	\begin{align}
	\mathcal{I}_{\Delta_0,\Delta_1, \Delta_2}(a, q_1,q_2)=\pi^{d/2}2^{\Sigma\Delta}\,\Gamma\left(\frac{\Sigma \Delta-d}{2}\right)\prod_{k=0}^2\int_{0}^{\infty} \frac{d t_k}{t_k}\,\frac{t_k^{\Delta_k}}{\Gamma(\Delta_k)}\,e^{T^2}\,.  
	\end{align}
	Plugging the expression $T^2=t_0^2\,X^2+2t_0t_1\,X\cdot q_1+2t_0t_2\,X\cdot q_1+2t_1t_2\,q_1\cdot q_2$ in the last equation and using again the Mellin-Barnes representation of the exponential functions, we obtain
	\begin{align}   
	\mathcal{I}_{\Delta_0,\Delta_1, \Delta_2}(a, q_1,q_2)&=\pi^{d/2}2^{\Sigma\Delta}\,\Gamma\left(\frac{\Sigma \Delta-d}{2}\right)\prod_{k=0}^2\int_{0}^{+\infty} \frac{d t_k}{t_k}\,\frac{t_k^{\Delta_k}}{\Gamma(\Delta_k)}\int_{c_{00}-i\infty}^{c_{00}+i\infty}\frac{d s_{00}}{2\pi i}\, \Gamma(s_{00})\left(-t_0^2 X^2\right)^{-s_{00}}\notag\\
	\times & \int_{c_{01}-i\infty}^{c_{01}+i\infty}\frac{d s_{01}}{2\pi i}\, \Gamma(s_{01})\left(-2t_0t_1 X\cdot q_1\right)^{-s_{01}}\int_{c_{02}-i\infty}^{c_{02}+i\infty}\frac{d s_{02}}{2\pi i}\, \Gamma(s_{02})\left(-2t_0t_2 X\cdot q_2\right)^{-s_{02}}\notag\\
	\times &\int_{c_{12}-i\infty}^{c_{12}+i\infty}\frac{d s_{12}}{2\pi i}\, \Gamma(s_{12})\left(-2t_1t_2 q_2\cdot q_1\right)^{-s_{12}}\,.
	\end{align}
	We also have
	\begin{equation}
	\int_{0}^{\infty}\frac{d t}{t}\, t^{\alpha}=2\pi \delta(\alpha)\,, \qquad (\Re\E(\alpha)>0)\,.
	\end{equation}
	Therefore, each integral in $t_k$ gives a different delta function, which allows us to straightforwardly evaluate three out of the four remaining Mellin-Barnes integrals. 
	Setting $s_{01}=\Delta_1-s_{12}$, $s_{02}=\Delta_2-s_{12}$, and $s_{00}=s_{12}+\frac{1}{2}(u-\Delta_1-\Delta_2)$ yields
	\begin{align}
	\label{first option}
	&\mathcal{A}_{\Delta_1, \Delta_2, m}^{\text{conv}}(q_1, q_2; a)= \frac{\pi^{d/2}(-a^2)^{\frac{\Delta_1+\Delta_2}{2}}}{\Gamma(\Delta_1)\Gamma(\Delta_2)\left(-a\cdot q_1\right)^{\Delta_1}\left(-a\cdot q_2\right)^{\Delta_2}} \int_{-i\infty}^{+i\infty}\frac{d u}{2\pi i} \,\Gamma\left(\frac{u+\Delta_1+\Delta_2-d}{2}\right)\notag\\
	& \times \left(\frac{m\sqrt{-a^2}}{2}\right)^{-u}\int_{\frac{d}{2}-i\infty}^{\frac{d}{2}+i\infty}\frac{d s_{12}}{2\pi i}\, \Gamma(s_{12})\Gamma(\Delta_1-s_{12})\Gamma(\Delta_2-s_{12})\Gamma\left(s_{12}+\frac{u-\Delta_1-\Delta_2}{2}\right) z^{-s_{12}}\,,
	\end{align}
	with
	\begin{equation}
	\label{z def}
	z=\frac{1}{2}\frac{(-a^2)(-q_1\cdot q_2)}{\left(-a\cdot q_1\right)\left(-a\cdot q_2\right)}\ , 
	\end{equation}
	which is positive for $a^\mu$ timelike and future-directed, as assumed in our analysis. Solving first the integral in $u$ yields
	\begin{align}
	\mathcal{A}_{\Delta_1, \Delta_2, m}^{\text{conv}}&(q_1, q_2; a)= \frac{4(2\pi)^{d/2} \left(-ma\cdot q_1\right)^{\Delta_2-\frac{d}{2}}\left(-ma\cdot q_2\right)^{\Delta_1-\frac{d}{2}}}{\Gamma(\Delta_1)\Gamma(\Delta_2) (- q_1\cdot q_2)^{\Delta_1+\Delta_2-\frac{d}{2}}}\,\mathcal{M}^{(m)}_{\Delta_1, \Delta_2}(q_1, q_2; a)\,,
	\end{align}
	with
	\begin{align}
	\mathcal{M}^{(m)}_{\Delta_1, \Delta_2}(q_1, q_2; a)=&\int_{-i\infty}^{+i\infty}\frac{d s_{12}}{2\pi i}\,  \Gamma\left(\frac{d}{2}-\Delta_1+s_{12}\right)\Gamma\left(\frac{d}{2}-\Delta_2+s_{12}\right)\Gamma\left(\Delta_1+\Delta_2-\frac{d}{2}-s_{12}\right)\notag\\
	&\times\left(\frac{(-q_1\cdot q_2)}{\left(-ma\cdot q_1\right)\left(-ma\cdot q_2\right)}\right)^{s_{12}} \left(m\sqrt{-a^2}\right)^{s_{12}}K_{s_{12}}\left(m\sqrt{-a^2}\right)\,.
	\end{align}
	This last integral representation is particularly suitable for studying the distribution of poles.
	Alternatively, to obtain a closed-form expression for \eqref{3MixedConv}, it is more convenient to first perform the integral in $s_{12}$. Integrals of this form have been extensively studied in the Appendix of \cite{Sleight:2019mgd}. Specifically, our integral can be recast as follows, 
	\begin{multline}
	\label{ints}
	\int_{\frac{d}{2}-i\infty}^{\frac{d}{2}+i\infty}\frac{d s_{12}}{2\pi i}\, \Gamma(s_{12})\Gamma(\Delta_1-s_{12})\Gamma(\Delta_2-s_{12})\Gamma\left(s_{12}+\frac{1}{2}(u-\Delta_1-\Delta_2)\right)\, z^{-s_{12}}=\\    \zeta^{d}\,\frac{\Gamma(\Delta_1)\Gamma(\Delta_2)\Gamma\left(\frac{u+\Delta_1-\Delta_2}{2}\right)\Gamma\left(\frac{u-\Delta_1+\Delta_2}{2}\right)}{\Gamma\left(\frac{u+\Delta_1+\Delta_2}{2}\right)}\,{}_2F_1\left(\Delta_1, \Delta_2, \frac{u+\Delta_1+\Delta_2}{2}; 1-\zeta^2\right)\,,
	\end{multline}
	where $\zeta=\sqrt{z}$, for $z<1$, and $\zeta=1/\sqrt{z}$, for $z>1$, with $z$ given by \eqref{z def}. For $z=1$, \eqref{ints} reduces to Barnes' First Lemma, ensuring the analyticity of the solution for any $z>0$. This yields
	\begin{align}
	\label{second option}
	&\mathcal{A}_{\Delta_1, \Delta_2, m}^{conv}(q_1, q_2; a)= \frac{\pi^{d/2}(-a^2)^{\frac{\Delta_1+\Delta_2}{2}}}{\left(-2a\cdot q_1\right)^{\Delta_1}\left(-2a\cdot q_2\right)^{\Delta_2}}\,\mathscr{B}_{\Delta_1, \Delta_2, m}(q_1,q_2; a)\,,
	\end{align}
	where 
	\begin{multline}
	\mathscr{B}_{\Delta_1, \Delta_2, m}(q_1, q_2; a)= \zeta^{d}\,\int_{-i\infty}^{+i\infty}\frac{d u}{2\pi i} \,\left(\frac{m\sqrt{-a^2}}{2}\right)^{-u} \frac{\Gamma\left(\frac{u+\Delta_1+\Delta_2-d}{2}\right)\Gamma\left(\frac{u+\Delta_1-\Delta_2}{2}\right)\Gamma\left(\frac{u-\Delta_1+\Delta_2}{2}\right)}{\Gamma\left(\frac{u+\Delta_1+\Delta_2}{2}\right)}\\
	\times {}_2F_1\left(\Delta_1, \Delta_2, \frac{u+\Delta_1+\Delta_2}{2}; 1-\zeta^2\right) \,. 
	\end{multline}
	This last integral can be explicitly solved by rewriting 
	\begin{equation}
	\begin{split}
	&{}_2F_1\left(\Delta_1, \Delta_2;\frac{u+\Delta_1+\Delta_2}{2}, 1-\zeta^2 \right)\\
	&=(\zeta^{2})^{\frac{u-\Delta_1-\Delta_2}{2}} {}_2F_1\left(\frac{u-\Delta_1+\Delta_2}{2}, \frac{u+\Delta_1-\Delta_2}{2};\frac{u+\Delta_1+\Delta_2}{2}, 1-\zeta^2 \right)\,,
	\end{split}
	\end{equation}
	and then using the \textit{Mixed Integral Theorem} on page 143 of \cite{Slater}. This theorem provides a solution in terms of a linear combination of Appell-like series, ensuring the convergence of \eqref{3MixedConv}. 
	Although the expression is quite cumbersome, the convergence conditions of the \textit{Mixed Integral Theorem} still hold true when the solution is analytically continued to $m\to \E^{i\left(\frac{\pi}{2}-\epsilon\right)}\,m$. At the final point of this analytic continuation, we recover the closed-form expression of \eqref{3Mixed} for $a^\mu$ timelike and future-directed. This last result can then be analytically extended to the whole Minkowski spacetime using suitable transformations, as in \cite{Iacobacci:2022yjo}, where the authors split Minkowski into four regions -- two inside and two outside the lightcone -- and used analytical transformations to connect points across them.  In this procedure, all analytic continuations, which ultimately correspond to rotations in the complex plane in a suitable set of coordinates, are uniquely fixed by the requirement of not crossing any branch cut of~\eqref{3Mixed}.

	\bibliography{bibl}
	\bibliographystyle{JHEP}  
\end{document}